\documentclass[11pt,a4paper,amssymb,amsmath, tightenlines]{article}

\usepackage{amsmath, amssymb, amsthm, latexsym, mathrsfs, url, color, epsfig, graphics, float, setspace}
\usepackage{sectsty}
\usepackage[T1]{fontenc}
\usepackage[bitstream-charter]{mathdesign}

\usepackage[pdftex]{hyperref} 

\newtheorem{definition}{Definition}[section]
\newtheorem{remark}{Remark}
\newtheorem{proposition}{Proposition}[section]
\newtheorem{lemma}{Lemma}[section]

\numberwithin{equation}{section}

\newcommand{\e}{{\rm e}}
\newcommand{\E}{\mathbb{E}}

\newcommand{\F}{\mathcal{F}}

\newcommand{\PR}{\mathbb{P}}
\newcommand{\M}{\mathbb{M}}
\newcommand{\mL}{\mathbb{L}}
\newcommand{\Q}{\mathbb{Q}}
\newcommand{\R}{\mathbb{R}}

\newcommand{\rd}{\textup{d}}

\newcommand{\indi}[1]{1\hspace{-.09cm}\textup{\textrm{l}}}

\newcommand{\nn}{\nonumber}

\begin{document}
\title{\bf Heat Kernel Framework for \\ Asset Pricing in Finite Time}
\author{Andrea Macrina \\ \\ {Department of Mathematics, University College London} \\ {London WC1E 6BT, United Kingdom} 
}
\date{\today}
\maketitle
\vspace{-0.75cm}
\begin{abstract}
\noindent
A heat kernel approach is proposed for the development of a flexible and mathematically tractable asset pricing framework in finite time. The pricing kernel, giving rise to the price system in an incomplete market, is modelled by weighted heat  kernels which are driven by multivariate Markov processes and which provide enough degrees of freedom in order to calibrate to relevant data, e. g. to the term structure of bond prices. It is shown how, for a class of models, the prices of bonds, caplets, and swaptions can be computed in closed form. The dynamical equations for the price processes are derived, and explicit formulae are obtained for the short rate of interest, the risk premium, and for the stochastic volatility of prices. Several of the closed-form asset price models presented in this paper are driven by combinations of Markovian jump processes with different probability laws. Such models provide a rich basis for consistent applications in several sectors of a financial market including equity, fixed-income, commodities, and insurance. The flexible, multidimensional and multivariate structure, on which the asset price models are constructed, lends itself well to the transparent modelling of dependence across asset classes. As an illustration, the impact on prices by spiralling debt, a typical feature of a financial crisis, is modelled explicitly, and contagion effects are readily observed in the dynamics of asset returns. 
\\\vspace{-0.2cm}
\\\vspace{-0.2cm}

\vspace{.75cm}
\noindent
{Introduction \hfill 2}\\
{Pricing kernel models and the pricing of bonds, caplets, and swaptions \hfill 4}\\
{Closed-form and explicit price models \hfill 7}\\
{Dynamical equations \hfill 12}\\
{Incomplete market models driven by LRBs \hfill 14}\\
{General asset pricing in finite time \hfill 20}\\
{Spiralling debt and its impact on international bond markets \hfill 26}\\
{Conclusions \hfill 31}\\
{References \hfill 32}
\end{abstract}
\section{Introduction}
In this paper, we shall take the view that in a modern asset pricing framework (i) pricing models should be coherent across all asset classes traded in a financial market, (ii) securities pricing used in the front offices of financial firms should be compatible with asset risk management, and (iii) pricing formulae should be applicable in the banking industry and also in the insurance sector. Expressed in other words, these three requirements state that modern pricing models ought to be consistent under the real probability measure $\PR$ and the risk-neutral measure $\Q$, and at the same time they should retain a high degree of flexibility and mathematical ease while guaranteeing the coherence of the price system for all financial assets.

In what follows, we propose an asset pricing framework that can be applied, in principle, to all asset classes and that is mathematically tractable so that Monte Carlo techniques are not necessary for scenario simulations of asset price dynamics. The proposed approach includes partial automatic calibration to market data such as initial prices of assets. The price system of assets traded in a financial market shall be developed by modelling the pricing kernel (state-price density) first. Once the stochastic framework for the pricing kernel is built and the connection with bond prices is established, we go on to show how price processes for other asset classes can be derived in a natural way. We also consider how the situation, in which the debt of a sovereign country gets out of control, can be incorporated in the same pricing framework without introducing extra assumptions to include effects of credit risk.  

The general setup of the asset pricing framework is developed in finite time, $t\in[0,U]$ for $U<\infty$. We model a financial market by a filtered complete probability space $(\Omega,\F,\PR,\{\F_t\})$ that satisfies the usual hypotheses (Protter \cite{Pro}), where $\PR$ denotes the real probability measure and $\{\F_t\}$ is the market filtration. We consider a (multi-dimensional) process $\{X_t\}$ on $(\Omega,\F,\PR)$ and assume that the market filtration is generated by $\{X_t\}$. We also assume that $\{X_t\}$ has the Markov property with respect to $\{\F_t\}$, its natural filtration. We introduce the pricing kernel process $\{\pi_t\}$ to model the market agent's preferences and the dynamics of interest rates in the economy which $\{\pi_t\}$ is associated with. We write $\{S_t\}_{0\le t\le T<U}$ for the price process of a dividend-paying asset, and let $\{D_t\}_{0\le t\le T<U}$ denote the (continuous) dividend stream up until $T$. Then the price $S_t$ at time $t$ is given by
\begin{equation}\label{pf}
 S_{t}=\frac{1}{\pi_t}\,\E^{\PR}\left[\pi_T S_T+\int_t^T \pi_u D_u \rd u\,\bigg\vert\,\F_t\right].
\end{equation}
In order to calculate asset prices explicitly, the following ingredients need to be specified: (i) The Markov process $\{X_t\}$ that generates the market filtration, and thus the market information; (ii) the pricing kernel $\{\pi_t\}$, and thus the dynamics of the interest rates and the agent's preferences; (iii) the random variable $S_T$ and the process $\{D_t\}$, thus the asset's terminal cash flow and the dividend stream, respectively. All ingredients are specified in such a way that the price process $\{S_t\}$ is adapted to the market filtration generated by $\{X_t\}$. For in-depth accounts about the theory of pricing kernels, preferences, asset pricing, and interest rates modelling, one may consult the textbooks by, e. g., Back \cite{back}, Bj\"ork \cite{tb}, Cochrane \cite{coch}, Duffie \cite{duf}, and by Brigo \& Mercurio \cite{bm}.

In the next section, we introduce weighted heat kernels to define the class of pricing kernels, and thus give rise to the asset pricing framework treated in this paper. Heat kernel models for the development of stochastic price systems have been proposed by Akahori {\it et al}. \cite{ahtt} in an infinite-time setting, and more recently by Akahori \& Macrina \cite{am} in a finite-time context. 

At first, we summarise the construction of the pricing kernel presented in Akahori \& Macrina \cite{am}, and at the same time, we extend the approach so that automatic partial calibration can be accommodated. Then we write the pricing kernel models and the resulting discount bond price processes in a concise formalism, which we show remains unchanged if one were to apply a different probability measure. We provide general formulae for the price processes of discount bonds, caplets and swaptions, and the associated nominal interest rate process. The stochastic short rate of interest is by construction non-negative. 

In Section \ref{Sec-cfmodels}, we show how the introduced formalism reveals a class of rational asset price models of which structure can be decomposed in a deterministic part and a martingale under an auxiliary measure. Thus we derive closed-form price processes for bonds, caplets, and swaptions, and explicit price models are then obtained by specifying the dynamics of the market information flow and the degrees of freedom in the formulation of a particular pricing kernel model. The dynamics of certain price processes have time-dependent lower and upper bounds, a feature we not necessarily view as being a shortcoming.  

In Section \ref{Sec-DynEq}, we derive the dynamical equation for the bond price processes introduced in the previous section for the case that the market filtration is generated by a continuous process. The market price of risk process is also obtained endogenously, which, for this class of pricing models, incorporates a discernible part that can be identified as the incentive for accepting model risk. 

In Section \ref{Sec-incompmarket}, we introduce multivariate L\'evy random bridges and extend the pricing framework to an incomplete market. A generalised class of asset price models characterised by a Fourier transform is presented, and asset price models with higher-order rational structures are developed. Explicit price processes driven by jump processes are derived.

In Section \ref{Gen-asset-p}, the proposed pricing kernel approach is applied to general asset pricing, and we show how asset price models constructed under the $\PR$-measure translate into asset price models equipped with stochastic interest rates and stochastic volatility under the $\Q$-measure. The interaction between the bond price process and, e. g., the equity component of the discounted share price process is clearly identifiable. This property renders the herewith proposed asset pricing framework also appealing for the construction of hedging strategies against losses due to the exposure of a financial position to a specific market sector. Additional examples of explicit multivariate asset pricing models driven by jump processes are given.  

In Section \ref{defic-mod}, spiralling sovereign deficit is modelled and its impact on the price dynamics of sovereign bonds is shown. The flexibility of the considered heat kernel state-price density approach allows for the construction of explicit dependence models linking the price evolution of, e. g., bonds issued by several sovereign governments. Contagion effects arise endogenously, and the graphs illustrate the impact of dependent economies and markets on the price dynamics of assets.

We concluded with listing the novel contributions presented in this article and with a research outlook which proposes several extensions and applications of the asset pricing framework considered next.
\section{Pricing kernel models and the pricing of bonds, caplets and swaptions}\label{Sec-II}
The backbone of the pricing framework considered in this paper is the following class of pricing kernel models $\{\pi_t\}_{0\le t\le U}$:
\begin{equation}\label{genpk}
 \pi_t=f_0(t)+f_1(t)\int^{U-t}_0\E\left[F(t+u,X_{t+u})\,\vert\,X_t\right]w(t,u)\,\rd u,
\end{equation}
where $\{X_t\}$ is an unspecified Markov process defined for $t\in[0,U]$ such that $t+u\le U$. Furthermore, $f_0(t)$ and $f_1(t)$ are deterministic, positive, and non-increasing functions, $F(t,x)$ is a positive measurable function, and $w(t,u)$ is a deterministic, positive and measurable function that satisfies 
\begin{equation*}
 w(t,u-s) \le w(t-s,u)
\end{equation*}
 for $s\le t\wedge u$. Assuming that $F(t,x)$ and $w(t,u)$ are chosen such that the integral in (\ref{genpk}) is finite for all $t$, it can be proven that the considered pricing kernel processes are indeed supermartingales adapted to the filtration generated by $\{X_t\}$. We refer to Akahori \& Macrina \cite{am}, Section 2, for a proof that can be applied also in the present context. 

As a special case of the general pricing formula (\ref{pf}), the price process of a discount bond with maturity $T$, is given by
\begin{equation}\label{db-pf}
 P_{tT}=\frac{1}{\pi_t}\,\E\left[\pi_T\,\vert\,X_t\right]
\end{equation}
where $0\le t\le T<U$. We keep in mind that the market filtration is generated by the Markov process $\{X_t\}$, and thus it suffices to take the expectation conditional only on $X_t$. The conditional expectation of $\pi_T$ can be computed explicitly to obtain
\begin{equation*}
 \E\left[\pi_T\,\vert\,X_t\right]=f_0(T)+f_1(T)\int^{U-t}_{T-t}\E\left[F(t+u,X_{t+u})\,\vert\,X_t\right]\,w(T,u-T+t)\,\rd u,
\end{equation*}
where the tower property is invoked and a variable substitution is applied. We define
\begin{equation}\label{Y-def}
 Y_{tT}=\int^{U-t}_{T-t}\E\left[F(t+u,X_{t+u})\,\vert\,X_t\right]\,w(T,u-T+t)\,\rd u.
\end{equation}
The bond price process can then be written in the compact form
\begin{equation}\label{bondpp}
 P_{tT}=\frac{f_0(T)+f_1(T)\,Y_{tT}}{f_0(t)+f_1(t)\,Y_{tt}},
\end{equation}
and the initial term structure is given by
\begin{equation*}
 P_{0t}=\frac{f_0(t)+f_1(t)\,Y_{0t}}{f_0(0)+f_1(0)\,Y_{00}}.
\end{equation*}
We deduce that 
\begin{equation}\label{f_0-function}
 f_0(t)=P_{0t}\left[1+f_1(0)Y_{00}\right]-f_1(t)Y_{0t},
\end{equation}
where we may set $f_0(0)=1$ with no loss of generality. By inserting (\ref{f_0-function}) in (\ref{bondpp}), we obtain
\begin{equation}\label{P-bond}
 P_{tT}=\frac{P_{0T}+y(T)\left(Y_{tT}-Y_{0T}\right)}{P_{0t}+y(t)\left(Y_{tt}-Y_{0t}\right)},
\end{equation}
where
\begin{equation*}
 y(t)=\frac{f_1(t)}{1+f_1(0)Y_{00}},
\end{equation*}
for $0\le t\le T$. Similarly, the expression for the pricing kernel can be written in the form
\begin{equation}\label{pricingkernel}
 \pi_t=\pi_0\left[P_{0t}+y(t)\left(Y_{tt}-Y_{0t}\right)\right],
\end{equation}
where $\pi_0=1+f_1(0)Y_{00}$. We note that the pricing kernel is now calibrated to the initial term structure of discount bonds. Assuming that the bond price function is differentiable with respect to $T$, the expression for the instantaneous forward rate $\{r_{tT}\}$ is given by
\begin{eqnarray}\label{f-rate}
 r_{tT}&=&-\partial_T\,\ln\left(P_{tT}\right),\nn\\
       &=&-\frac{\partial_T P_{0T}+\left(Y_{tT}-Y_{0T}\right)\partial_T y(T)+y(T)\left(\partial_T Y_{tT}-\partial_T Y_{0T}\right)}{P_{0T}+y(T)\left(Y_{tT}-Y_{0T}\right)}.
\end{eqnarray}
 The process $\{r_t\}$ for the short rate of interest can then be deduced by setting $r_t={r_{tT}}_{\big\vert{T=t}}$:
\begin{equation}\label{SRI}
 r_t=-\frac{\partial_t P_{0t}+\left(Y_{tt}-Y_{0t}\right)\partial_t y(t)+y(t)\left\{\left(\partial_T Y_{tT}-\partial_T Y_{0T}\right)\right\}_{T=t}}{P_{0t}+y(t)\left(Y_{tt}-Y_{0t}\right)}.
\end{equation}

Once the bond price system is derived, one can calculate the price of fixed-income derivatives such as caplets and swaptions. We consider a $t$-maturity swaption contract with strike $K$, which is written on a collection of discount bonds $P_{tT_i}$ with maturities $\{T_i\}_{1,\ldots,n}$. An application of the pricing formula (\ref{pf}) shows that the price $Swp_{0t}$ of the swaption at time zero is 
\begin{equation}\label{Swaption}
 Swp_{0t}=\frac{1}{\pi_0}\E\left[\pi_t\left(1-P_{tT_n}-K\sum^n_{i=1}P_{tT_i}\right)^+\right],
\end{equation}
where 
\begin{equation*}
 P_{tT_i}=\frac{P_{0T_i}+y(T_i)\left(Y_{tT_i}-Y_{0T_i}\right)}{P_{0t}+y(t)\left(Y_{tt}-Y_{0t}\right)}.
\end{equation*}
Here, for simplicity, we consider a swaption written on a payer's interest rate swap with unit notional and unit year fraction. The maturity of the swaption is assumed to coincide with the first of the $n$ reset dates of the underlying swap, that is $t=T_0$. Then, by use of (\ref{pricingkernel}), we obtain
\begin{eqnarray*}
 Swp_{0t}&=&\E\left[\left(P_{0t}-P_{0T_n}-K\sum^n_i P_{0T_i}+y(t)\left(Y_{tt}-Y_{0t}\right)\right.\right.\\ \nn
&&\hspace{1cm}\left.\left.-y(T_n)\left(Y_{tT_n}-Y_{0T_n}\right)-K\sum^n_{i=1}y(T_i)\left(Y_{tT_i}-Y_{0T_i}\right)\right)^+\right].
\end{eqnarray*}
The price of caplets can be calculated in an analogous way. Further details for the calculation of caplets and swaptions prices follow in the next section.

We may wonder at this stage whether it might be possible to construct a class of discount bond price processes for which the associated prices of interest rate derivatives can be calculated in closed form. We shall present such bond price models in detail in the next section. Before, we prepare the ground by making the following observation.
\begin{proposition}\label{Prop-M-pk}
 Let $\{M_t\}_{0\le t<U}$ be an $\{\F_t\}$-adapted $\PR$-martingale that induces a change-of-measure from $\PR$ to an equivalent auxiliary probability measure $\M$ with respect to which the Markov property of $\{X_t\}$ is preserved. Let (i) $f_0(t)$ and $f_1(t)$ be deterministic, positive and non-increasing functions, (ii) $F(t,x)$ be a measurable positive function, and (iii) $w(t,u)$ be a positive measurable function satisfying
\begin{equation*}
 w(t,u-s) \le w(t-s,u),
\end{equation*}
 for $s\le t\wedge u$. Let $\{Y^{\M}_{tT}\}_{0\le t\le T<U}$ be defined by 
\begin{equation*}
 Y^{\M}_{tT}=\int^{U-t}_{T-t}\E^{\M}\left[F(t+u,X_{t+u})\,\vert\,X_t\right]\,w(T,u-T+t)\,\rd u,
\end{equation*}
where $F(t,x)$ and $w(t,u)$ are such that $Y^{\M}_{tT}<\infty$ for all $t$. 
Then, the process
\begin{equation}\label{M-pk}
 \pi_t=\frac{\pi_0}{M_0}\left[P_{0t}+y^{\M}(t)\left(Y^{\M}_{tt}-Y^{\M}_{0t}\right)\right]M_t
\end{equation}
is a positive $(\{\F_t\},\PR)$-supermartingale where $\pi_0=1+f_1(0)Y^{\M}_{00}$,
\begin{equation*}
 P_{0t}=\frac{f_0(t)+f_1(t)\,Y^{\M}_{0t}}{1+f_1(0)\,Y^{\M}_{00}},
\end{equation*}
and
\begin{equation*}
 y^{\M}(t)=\frac{f_1(t)}{1+f_1(0)Y^{\M}_{00}}.
\end{equation*}
Taking (\ref{M-pk}) as the model for the pricing kernel, it follows that the price process $\{P_{tT}\}$ of the discount bond is given by
 \begin{equation}\label{M-bond}
 P_{tT}=\frac{P_{0T}+y^{\M}(T)\left(Y^{\M}_{tT}-Y^{\M}_{0T}\right)}{P_{0t}+y^{\M}(t)\left(Y^{\M}_{tt}-Y^{\M}_{0t}\right)},
\end{equation}  
where $P_{0t}$ is the initial term structure for $0\le t\le T<U$.
\end{proposition}

Proof. By inserting the expressions for $\pi_0$, $P_{0t}$ and $y^{\M}(t)$ in the model (\ref{M-pk}), we recover $\pi_t=M_t/M_0[f_0(t)+f_1(t)Y^{\M}_{tt}]$. Then, for $0\le s\le t<U$, it follows that
\begin{eqnarray}
 \E^{\PR}\left[\pi_t\,\vert\,\F_s\right]&=&\E^{\PR}\left[\frac{M_t}{M_0}\left(f_0(t)+f_1(t)Y^{\M}_{tt}\right)\,\big\vert\,\F_s\right],\nn\\
 							    &=&\frac{M_s}{M_0}\,\E^{\M}\left[f_0(t)+f_1(t)Y^{\M}_{tt}\,\big\vert\,\F_s\right],\nn\\
							    &\le&\frac{M_s}{M_0}\,\E^{\M}\left[f_0(s)+f_1(s)Y^{\M}_{ss}\right]=\pi_s
\end{eqnarray}
since the process $\{f_0(t)+f_1(t)Y^{\M}_{tt}\}$ is an $(\{\F_t\},\M)$-supermartingale; compare with (\ref{genpk}) via (\ref{Y-def}). To verify the validity of (\ref{M-bond}), the pricing kernel model (\ref{M-pk}) is inserted in (\ref{db-pf}) and the abstract version of the Bayes formula is applied to obtain the discount bond price processes expressed under the $\M$-measure.
\hfill$\Box$

Thus, if convenient, we can construct pricing kernel models by making use of the $\M$-measure while using the form (\ref{M-pk}). By comparing (\ref{P-bond}) with (\ref{M-bond}), we observe that the form of the asset price formulae remains unchanged.
\section{Closed-form and explicit price models}\label{Sec-cfmodels}
In this section, we construct a class of pricing kernels for which the price processes of underlying and derivative assets are obtained analytically. We explicitly calculate the price processes of bonds, caplets, and swaptions, and note that some models lead to bounded price processes---a property we not necessarily view as a shortcoming.\\   
In order to obtain explicit pricing models, the following quantities need to be specified in the definition of the pricing kernel (\ref{genpk}): (i) The finite-time Markov process $\{X_t\}$ that generates the market filtration and drives all prices, (ii) the positive function $F(t+u,x)$ that, to a great extent, characterises the type of pricing model (the DNA of the model), (iii) the weight function $w(t,u)$, and (iv) the deterministic functions $f_0(t)$ and $f_1(t)$. The particular class of models considered in this paper allows for explicit calibration of $f_0(t)$ to the initial term structure, and for a one-to-one correspondence between the degree of freedom $f_1(t)$ and option data (e. g., caplets and swaptions). Thus we only specify the Markov process $\{X_t\}$, $F(t+u,x)$, and $w(t,u)$.
\\

\noindent{\bf Quadratic and exponential quadratic models}. Two examples are provided in Akahori \& Macrina \cite{am}, namely the {\it quadratic} and the {\it exponential quadratic classes}. Here, for example, the Markov process $\{X_t\}$, that generates $\{\F_t\}$, is taken to be the Brownian random bridge $\{L_{tU}\}$ defined by
\begin{equation}\label{BRB}
 L_{tU}=\sigma\, t\, X_U + \beta_{tU},
\end{equation}
where $\sigma$ is a constant parameter, $X_U$ is an $\F_U$-measurable random variable with {\it a priori} density $p(x)$, and $\{\beta_{tU}\}_{0\le t\le U}$ is an independent standard Brownian bridge where $U$ is fixed. In order to obtain the quadratic models, $F(t+u,x)=x^2$ and $w(t,u)=U-t-u$ are chosen and inserted in the pricing kernel model (\ref{genpk}). By following the steps from (\ref{db-pf}) to (\ref{bondpp}) one arrives at the explicit expresssion for the discount bond price process within the quadratic class. This is one way. Another way is to realise that there exists a probability measure $\M$, equivalent to $\PR$, under which the Markov process (\ref{BRB}) has the law of a standard Brownian bridge for $t\in[0,U)$. The change of measure from $\PR$ to $\M$ is induced by the density martingale $\{M_t\}_{0\le t<U}$ defined by
\begin{equation*}
 \rd M_t=-\,\frac{\sigma U}{U-t}\E^{\PR}\left[X_U\,\vert\,L_{tU}\right]M_t\,\rd W^{\PR}_t
\end{equation*}
where $\{W^{\PR}_t\}$ is a Brownian motion process specified by (\ref{W-inno}) in the next section. The expectation $\E^{\PR}\left[X_U\,\vert\,L_{tU}\right]$ can be calculated by the conditional version of Bayes formula as shown for instance in Brody {\it et al}. \cite{bhm}. Thus, instead of deriving the explicit model for $\{P_{tT}\}$ by use of the $\PR$-measure, we instead apply Proposition \ref{Prop-M-pk}, in particular (\ref{M-bond}), and work with the probability measure $\M$, which simplifies the task of working out conditional expectations. As we shall see in Section \ref{Sec-incompmarket}, changing the measure to an auxiliary probability measures has also other advantages. 
The expression (\ref{M-bond}) for the discount bond then simplifies such that it can be written in the form
\begin{equation}\label{A-db}
 P_{tT}=\frac{P_{0T}+b(T)\,A_t}{P_{0t}+b(t)\,A_t}
\end{equation}
where, for the quadratic class, one has
\begin{align}\label{Quad-Ab}
 &P_{0t}=\frac{f_0(t)+\frac{1}{12\,U}(U-t)^3(U+3t)\,f_1(t)}{1+\frac{1}{12}U^3\,f_1(0)},& &b(t)=\frac{(U-t)^4\,f_1(t)}{4U\left[1+\tfrac{1}{12}U^3\,f_1(0)\right]},&\nn\\
 &A_t=\frac{U}{(U-t)^2}\,L_{tU}^2-\frac{t}{U-t},& &0\le t\le T<U.&
\end{align} 
An analogous calculation leads to the explicit expression for the discount bond price process within the exponential quadratic class which is characterised by
\begin{equation*}
F(t,x)=\exp\left(\frac{x^2}{2\,(U-t)}\right)
\end{equation*}
and $w(t,u)=(U-t-u)^{\eta-1/2}$, $\eta>1/2$.
The explicit bond price process (\ref{bondpp}), or equivalently (\ref{M-bond}),  can then also be written in the form (\ref{A-db}) where
\begin{align}\label{ExpQuad-Ab}
 &P_{0t}=\frac{f_0(t)+\frac{1}{\eta}(U-t)^{\eta}U^{1/2}f_1(t)}{1+\frac{1}{\eta}U^{\eta+1/2}f_1(0)},& &b(t)=\frac{\frac{1}{\eta}(U-t)^{\eta}U^{1/2}f_1(t)}{1+\frac{1}{\eta}U^{\eta+1/2}f_1(0)},&\nn\\
 &A_t=\sqrt{1-\frac{t}{U}}\,\exp\left(\frac{L_{tU}^2}{2(U-t)}\right)-1,& &0\le t\le T<U.&
\end{align}
It can be proven that the process $\{A_t\}_{0\le t<U}$, in (\ref{Quad-Ab}) and in (\ref{ExpQuad-Ab}), is an $(\{\F_t\},\M)$-martingale by showing $\E^{\M}[\vert M_t\vert]<\infty$ and $\E^{\M}[A_t\vert\F_t]=A_s$ for $0\le s\le t<U$. This property makes the representation (\ref{A-db}) appealing for the construction of further models. We summarise the various observation as follows:
\begin{lemma}\label{bA-models}
Given the pricing kernel class (\ref{genpk}), there exist equivalent measures $\M$, deterministic functions $b(t)$ and $(\{\F_t\},\M)$-martingales $\{A_t\}_{0\le t<U}$ such that
\begin{equation}\label{Ab-pk}
\pi_t=\frac{\pi_0}{M_0}\left[P_{0t}+b(t)\,A_t\right]M_t
\end{equation}
where $\{M_t\}_{0\le t<U}$ is the density martingale inducing the change-of-measure $\PR\rightarrow\M$, and $P_{0t}$ is the initial term structure of the discount bond system with price process 
\begin{equation}\label{Ab-bond}
P_{tT}=\frac{P_{0T}+b(T)\,A_t}{P_{0t}+b(t)\,A_t}
\end{equation}
for $0\le t\le T<U$. Assuming that the discount bond system is differentiable with respect to its maturity parameter $T$, the associated short rate of interest process $\{r_t\}$ is given by
\begin{equation}\label{Ab-rate}
 r_t=-\frac{\partial_t P_{0t}+A_t\partial_t b(t)}{P_{0t}+b(t)\,A_t}.
\end{equation}
\end{lemma}

Proof. To prove (\ref{Ab-rate}), we first derive the instantaneous forward rate $\{r_{tT}\}$ as in (\ref{f-rate}), that is by $r_{tT}=-\partial_T\ln\left(P_{tT}\right)$, and then we set $T=t$ in the resulting process $\{r_{tT}\}$ which gives us the short rate $\{r_t\}$. \hfill$\Box$
\\

Although written in the less unifying form proposed in Lemma \ref{bA-models}, and in the case of the quadratic models with fewer useful degrees of freedom, the quadratic and exponential quadratic models were first developed in Akahori \& Macrina \cite{am}. The exponential quadratic class appears also in Hughston \& Macrina \cite{hm}. We also note that interest rate models with a quadratic or exponential quadratic structure have been studied in Jamashidian \cite{jam} and McCloud \cite{McC1,McC2}, too. More examples of asset price models of the form (\ref{Ab-bond}), and thus associated with a pricing kernel of the type (\ref{Ab-pk}), are given in Sections \ref{Sec-incompmarket}, \ref{Gen-asset-p} and \ref{defic-mod}. Those models also include jumps in their dynamics.

\paragraph{\bf Caplets and swaptions.} We now proceed to calculate the prices of caplets and swaptions when the pricing kernel model is of the form (\ref{Ab-pk}). At time 0, the price $C_{0t}$ of a caplet with maturity $t$ and strike $K$ can be expressed as the price of a put bond option, that is
\begin{equation}
C_{0t}=\frac{1}{\pi_0}\E^{\PR}\left[\pi_t\left(K-P_{tT}\right)^+\right].
\end{equation}  
Expressions (\ref{Ab-pk}) and (\ref{Ab-bond}) are plugged in, whereby $M_t/M_0$ in (\ref{Ab-pk}) is utilised to change the probability measure from $\PR$ to $\M$. Since $P_{0t}+b(t)\,A_t$ is by construction positive, it can be taken inside the $max$-function so to obtain 
\begin{equation}
C_{0t}=\E^{\M}\left[\left(KP_{0t}-P_{0T}+\left[Kb(t)-b(T)\right]A_t\right)^+\right].
\end{equation}
In order for the caplet to be in-the-money, $[Kb(t)-b(T)]\,A_t>P_{0T}-KP_{0t}$ needs to hold. There are three cases: (i) $Kb(t)-b(T)>0$ leading to $A_t>(P_{0T}-KP_{0t})/\vert Kb(t)-b(T)\vert$, (ii) $Kb(t)-b(T)<0$ leading to $A_t<-(P_{0T}-KP_{0t})/\vert Kb(t)-b(T)\vert$, and (iii) $Kb(t)-b(T)=0$ resulting in $A_t$ being undetermined, in which case we may set $C_{0t}=0$. The caplet price can thus be written as follows:
\begin{equation}\label{caplet-int}
C_{0t}=\left(KP_{0t}-P_{0T}\right)\int_{a^{\ast}}p(a)\rd a+\left[Kb(t)-b(T)\right]\int_{a^{\ast}}a\,p(a)\rd a
\end{equation}
where $p(a)\rd a=\M[A_t\in\rd a]$. Furthermore, $a^{\ast}=\{a:\,a>a_c\ \textrm{if}\ Kb(t)-b(T)>0\}$ or $a^{\ast}=\{a:\,a<-a_c\ \textrm{if}\ Kb(t)-b(T)<0\}$, where
\begin{equation}
a_c:=\frac{P_{0T}-KP_{0t}}{\vert Kb(t)-b(T)\vert}.
\end{equation}
We set $C_{0t}=0$ for the case $Kb(t)-b(T)=0$. As an example, we apply the caplet price formula (\ref{caplet-int}) in the case that the pricing kernel is modelled by the quadratic class (\ref{Quad-Ab}). The $\M$-density $p(a)$ is quadratic Gaussian in this class, and the critical value $\kappa$ necessary to calculate the value of the in-the-money option is
\begin{equation}
\kappa=\sqrt{1+\frac{U-t}{t}\,\frac{P_{0t}-KP_{0T}}{\vert\,Kb(t)-b(T)\,\vert}}.
\end{equation}
Then, the caplet price, in the quadratic class, can be expressed as follows: For $K\,b(t)-b(T)>0$, we have
\begin{equation}
C_{0t}=2\left(KP_{0t}-P_{0T}\right)N(-\kappa)+\sqrt{\frac{2}{\pi}}\frac{\vert\,Kb(t)-b(T)\vert\,t}{U-t}\,\kappa\,\exp\left(-\tfrac{1}{2}\kappa^2\right).
\end{equation}
where $N(x)$ denotes the cumulative normal distribution function. For $K\,b(t)-b(T)<0$, the caplet price is
\begin{equation}
C_{0t}=\left(KP_{0t}-P_{0T}\right)\left(2\,N(\kappa)-1\right)+\sqrt{\frac{2}{\pi}}\,\frac{\vert\,K\,b(t)-b(T)\vert\,t}{U-t}\,\kappa\,\exp\left(-\tfrac{1}{2}\kappa^2\right).
\end{equation}
For $K\,b(t)-b(T)=0$, we set $C_{0t}=0$, by definition. 

A similar calculation leads to the analytical caplet price in the case one applies the exponential quadratic class (\ref{ExpQuad-Ab}). We shall give the details of an example utilising the exponential quadratic class when computing the explicit price of a swaption.

We recall that the price $Swp_{0t}$ of a swaption with maturity $t$ and strike $K$ is given by (\ref{Swaption}). Assuming that the pricing kernel model is of the form (\ref{Ab-pk}), we obtain
\begin{equation}
Swp_{0t}=\E^{\M}\left[\left(P_{0t}-P_{0T_n}-K\sum^n_{i=1}P_{0T_i}+\left[b(t)-b(T_n)-K\sum^n_{i=1}b(T_i)\right]A_t\right)^+\right].
\end{equation}
The swaption price is non-zero if the argument of the $max$-function is positive.  We are lead to three cases that depend on the sign of $b(t)-b(T_n)-K[b(T_1)+\cdots +b(T_n)]$. Similar to the calculation of the caplet price above, the swaption price can be written in a form that includes the three cases:
\begin{eqnarray}
Swp_{0t}&=&\left(P_{0t}-P_{0T_n}-K\sum^n_{i=1}P_{0T_i}\right)\int_{a^{\ast}}p(a)\rd a\nn\\
		  &&\hspace{2cm}+\left[b(t)-b(T_n)-K\sum^n_{i=1}b(T_i)\right]\int_{a^{\ast}}a\,p(a)\rd a
\end{eqnarray} 
where $p(a)\rd a=\M[A_t\in\rd a]$. Here, $a^{\ast}=\{a:\,a>a_S\ \textrm{if}\ b(t)-b(T_n)-K[b(T_1)+\cdots +b(T_n)]>0\}$ or $a^{\ast}=\{a:\,a<-a_S\ \textrm{if}\ b(t)-b(T_n)-K[b(T_1)+\cdots +b(T_n)]<0\}$ where
\begin{equation}
a_S:=\frac{K\sum^n_{i=1}P_{0T_i}+P_{0T_n}-P_{0t}}{\vert b(t)-b(T_n)-K\sum^n_{i=1}b(T_i)\vert}.
\end{equation}
In the case that $b(t)-b(T_n)-K[b(T_1)+\cdots +b(T_n)]=0$, we set $Swp_{0t}=0$. We note here that the strike $K$, the maturity $t$ and the reset dates $T_i$, $i=1,\ldots,n$, are all fixed contractually. This means that the discriminant as to which of the three cases prevails, is essentially predetermined in the swaption contract once a specific pricing model is chosen.

Next, we compute the price of a swaption by use of the exponential quadratic class (\ref{ExpQuad-Ab}). The swaption price is given as follows: We first define
\begin{equation}
\nu=\sqrt{\frac{2U}{t}\ln\left[\left(1-\frac{t}{U}\right)^{-1/2}\left(\frac{K\sum^n_{i=1}P_{0T_i}+P_{0T_n}-P_{0t}}{\vert b(t)-b(T_n)-K\sum^n_{i=1}b(T_i) \vert}+1\right)\right]}.
\end{equation}
Then, for $b(t)-b(T_n)-K[b(T_1)+\cdots +b(T_n)]>0$, we have
\begin{eqnarray}
Swp_{0t}&=&2\left(P_{0t}-P_{0T_n}-K\sum^n_{i=1}P_{0T_i}\right)N(-\nu)\nn\\
		   &+&2\left[b(t)-b(T_n)-K\sum^n_{i=1}b(T_i)\right]\left(N\left(-\sqrt{1-\frac{t}{U}}\ \nu\right)-N\left(-\nu\right)\right).
\end{eqnarray}
For $b(t)-b(T_n)-K[b(T_1)+\cdots +b(T_n)]<0$, we have
\begin{eqnarray}
Swp_{0t}&=&\left(P_{0t}-P_{0T_n}-K\sum^n_{i=1}P_{0T_i}\right)\left(2N(\nu)-1\right)\nn\\
		   &-&2\,\vert b(t)-b(T_n)-K\sum^n_{i=1}b(T_i) \vert\left(N\left(\sqrt{1-\frac{t}{U}}\ \nu\right)-N\left(\nu\right)\right).
\end{eqnarray}
For the case $b(t)-b(T_n)-K[b(T_1)+\cdots +b(T_n)]=0$, we have $Swp_{0t}=0$, by definition. An analogous calculation leads to the explicit price of a swaption when applying the quadratic models (\ref{Quad-Ab}).
\paragraph{\bf Boundedness of prices.} Bond prices fluctuate by construction between zero and one, and the associated interest rate is non-negative. However, the bond price processes produced by the above models, e. g. the quadratic and the exponential quadratic models, have tighter time-dependent bounds, and the same holds for the interest rate and the yield of the bond. One might think that having bounded bond prices and associated interest rates is a shortcoming. On the contrary, we think that such a feature may be advantageous, especially if the time-dependent bounds are wide enough for the interest rate to have sufficient freedom. The bounds may be put in relation with economic policies of which goal is to keep bond prices within a certain range. In turn, this may suggest to use the additional degree of freedom $f_1(t)$, cast inside the deterministic function $b(t)$, to include regulators' policies, for instance. Research regarding bounded asset prices and the relation to regulators' policies and markets might be continued elsewhere. We keep the boundedness property inherent in certain rational asset pricing models in our mind for when we later turn to general asset pricing, Section \ref{Gen-asset-p}, and to the impact on prices by an economy's spiralling deficit, Section \ref{defic-mod}. The higher-order price models presented in Section \ref{Sec-incompmarket} do not necessarily exhibit tighter bounds.   
\section{Dynamical equations}\label{Sec-DynEq} 
We derive the dynamical equation of the bond price for the case that the martingale $\{A_t\}$ introduced in Lemma \ref{bA-models} is an Ito process. The expression of the bond price dynamics reveals the market price of risk, which is obtained endogenously. It turns out that the market price of risk process is constituted by two distinct stochastic components. We highlight an example in which the unambiguous interpretation of the two components emerges naturally:  One part of the risk premium is associated with the stochasticity of a financial market due to noisy information about the market factors. The second part of the risk premium can be identified as model risk that is directly related to the choice of the class of price models. Furthermore, the Brownian motion that drives the bond price process arises also endogenously, and is identified with the innovations process updating the price process as the quality of market information improves.  
\begin{proposition}\label{Prop-db-SDE}
Let $\{W^{\PR}_t\}_{0\le t<U}$ be a standard $(\{\F_t\},\PR)$-Brownian motion. Let the $(\{\F_t\},\M)$-martingale $\{A_t\}_{0\le t<U}$, considered in Lemma \ref{bA-models},  satisfy
\begin{align}\label{A-Ito}
 &\rd A_t=\nu_t\left(\rd W^{\PR}_t+\vartheta_t\rd t\right)\nn\\
 &A_0=0,
\end{align}
where $\{\nu_t\}_{0\le t<U}$ and $\{\vartheta_t\}_{0\le t<U}$ are well-defined $\{\F_t\}$-adapted processes. Then, the discount bond price process (\ref{Ab-bond}) satisfies
\begin{equation}\label{Ab-bond-dyn}
 \frac{\rd P_{tT}}{P_{tT}}=\left(r_t+\lambda_t\Omega_{tT}\right)\rd t+\Omega_{tT}\rd W^{\PR}_t,
\end{equation}
where $\{r_t\}$ is determined by (\ref{Ab-rate}) and 
\begin{eqnarray}\label{rlO}
 \lambda_t&=&\vartheta_t-\nu_t\frac{b(t)}{P_{0t}+b(t)\,A_t},\label{A-IMPR}\\
 \Omega_{tT}&=&\nu_t\left[\frac{b(T)}{P_{0T}+b(T)\,A_t}-\frac{b(t)}{P_{0t}+b(t)\,A_t}\right].\nn
\end{eqnarray}
\end{proposition}
We note here that, via the Girsanov Theorem, we have introduced an $(\{\F_t\},\M)$-Brownian motion $\{W^{\M}_t\}$ that satisfies $\rd W^{\M}_t=\rd W^{\PR}_t+\vartheta_t\rd t$.
\begin{lemma}\label{Lemma-db-SDE}
The instantaneous forward rate $\{r_{tT}\}$ of a bond with price dynamics (\ref{Ab-bond-dyn}) satisfies the dynamical equation
\begin{equation}\label{A-IFR}
 \frac{\rd r_{tT}}{r_{tT}}=-\,\sigma_{tT}\,\Omega_{tT}\rd t+\sigma_{tT}\left(\rd W^{\PR}_t+\lambda_t\rd t\right),
\end{equation}
where the instantaneous forward rate volatility $\{\sigma_{tT}\}$ is defined by
\begin{equation}\label{Ab-fwd}
 \sigma_{tT}=-\,\nu_t\left(\frac{b(T)}{P_{0T}+b(T)\,A_t}-\frac{\partial_T b(T)}{\partial_T P_{0T}+A_t\partial_T b(T)}\right).
\end{equation}
\end{lemma}

The dynamical equations derived in Proposition \ref{Prop-db-SDE} and in Lemma \ref{Lemma-db-SDE} are obtained by applying Ito's differentiation rules to (\ref{A-db}) given the specification (\ref{A-Ito}).  The stochastic differential equation for the short rate of interest $\{r_t\}$ is given by
\begin{equation}\label{Ab-shortIR}
\frac{\rd r_t}{r_t}=\left(\frac{\partial_{tt}P_{0t}+A_t\partial_{tt}b(t)}{\partial_t P_{0t}+A_t\partial_t b(t)}+r_t\right)\rd t+\sigma_t\left(\rd W^{\PR}_t+\lambda_t\rd t\right),
\end{equation}
where $\{\lambda_t\}$ is the instantaneous market price of risk (\ref{A-IMPR}) and the volatility $\sigma_t$ is obtained by setting $T=t$ in (\ref{Ab-fwd}). Here we implicitly assume that the deterministic function $b(t)$ is twice differentiable. We emphasise that the deduced instantaneous forward rates have the HJM-form, c.f. Heath {\it et al.} \cite{hjm}, Filipovi\'c \cite{fil}. The SDEs (\ref{A-IFR}) and (\ref{Ab-shortIR}) can be written with respect to the risk-neutral measure $\Q$, which is associated with the market price of risk $\{\lambda_t\}$, by introducing and $(\{\F_t\},\Q)$-Brownian motion $\{W_t^{\Q}\}_{0\le t<U}$ that satisfies $\rd W^{\Q}_t=\rd W^{\PR}_t+\lambda_t\rd t$.
\paragraph{Model risk.} The risk premium $\{\lambda_t\}$ in (\ref{rlO}) is composed by two adapted processes, that is $\{\vartheta_t\}$ and another that incorporates $\{\nu_t\}$. In order to better understand the role these two components play, we consider the bond price processes generated by (\ref{Quad-Ab}) and (\ref{ExpQuad-Ab}) and which are driven by the Markov process (\ref{BRB}). The dynamical equation satisfied by these bond price models belongs to the class produced in Proposition \ref{Prop-db-SDE}, and the Brownian motion $\{W^{\PR}_t\}$ satisfies 
\begin{equation}\label{W-inno}
\rd W^{\PR}_t=\rd L_{tU}-\frac{1}{U-t}\left(\sigma U\,\E^{\PR}\left[X_U\,\vert\,L_{tU}\right]-L_{tU}\right)\rd t.
\end{equation}
It can be shown, by L\'evy's characterisation of Brownian motion, that the process satisfying (\ref{W-inno}) is indeed an $(\{\F_t\},\PR)$-Brownian motion for $t\in[0,U)$, see Brody {\it et al}. \cite{bhm}. Furthermore, it follows that the process $\{\vartheta_t\}$ is given by 
\begin{equation*}
 \vartheta_t=\frac{\sigma U}{U-t}\E^{\PR}\left[X_U\,\vert\,L_{tU}\right],
\end{equation*}
and that $\{\nu_t\}$ satisfies (i)
\begin{equation*}
 \nu_t=\frac{2U}{(U-t)^2}\,L_{tU}
\end{equation*}
in the case of the quadratic models (\ref{Quad-Ab}), and (ii)
\begin{equation*}
 \nu_t=\frac{L_{tU}}{\sqrt{U(U-t)}}\,\exp\left[\frac{L^2_{tU}}{2(U-t)}\right]
\end{equation*}
in the case of the exponential quadratic models (\ref{ExpQuad-Ab}). We thus observe that (i) $\{\vartheta_t\}$ is determined by the filtration model, that is, by the choice of the generating Markov process $\{L_{tU}\}$, and (ii) $\{\nu_t\}$ depends on the selection of the heat kernel models, that is, on the choice for $F(t,x)$ and $w(t,u)$. We can view $\{\vartheta_t\}$ as the risk premium component associated with the uncertainty in the market modelled via the information flow process $\{L_{tU}\}$. The component $\{\nu_t\}$ may however be interpreted as the premium associated with model risk since it is closely related to the choice of the specific asset price model. 
\section{Incomplete market models driven by LRBs}\label{Sec-incompmarket}
In this section, we extend the pricing framework to include multi-dimensional risk factors, and we generate asset pricing models in an incomplete market. We consider a class of finite-time Markov processes, the so-called ``L\'evy random bridges'' (LRBs), as constructed in Hoyle {\it et al}. \cite{hhm1}. An LRB can be interpreted as a L\'evy process that is bound to match a prescribed, albeit arbitrary, distribution at a fixed future time. The LRB and the generating L\'evy process are linked by an equivalent probability measure with respect to which the LRB has the law of the generating L\'evy process. Before we state the multi-variate version of this result appearing in Hoyle {\it et al}. \cite{hhm1}, we first give the definition of a multivariate LRB. This definition also establishes the notation of what follows in the subsequent sections.
\begin{definition}
 We say that $\{L_{tU}\}_{0\le t\le U}$ is a multivariate LRB on $\R^m$ if the following are satisfied:
\\ \\
1. The random variable $L_{UU}$ on $\R^m$ has marginal law $\nu$.\\ \\
2. There exist a multivariate L\'evy process $\{L_t\}_{0\le t\le U}$ on $\R^m$ such that $L_t$ has multivariate density function $\rho_t(x)$ on $\R^m$ for all $t\in(0,U]$.\\ \\
3. The marginal law $\nu$ concentrates mass where $\rho_U(z)$ is positive and finite, that is $0<\rho_U(z)<\infty$ for $\nu$-almost-every $z\in\R^m$. \\ \\
4. For every $n\in\mathbb{N}_+$, every $0<t_1<\ldots<t_n<U$, every $(x_1,\ldots,x_n)\in\R^m\times\R^n$, and $\nu$-almost-every $z\in\R^m$, we have
\begin{equation*}
\PR\left[L_{t_1 U}\le x_1,\ldots,L_{t_n U}\le x_n\,\vert\,L_{UU}=z\right]=\PR\left[L_{t_1}\le x_1,\ldots\,L_{t_n}\le x_n\,\vert\,L_U=z\right].
\end{equation*}
\end{definition}
\begin{proposition}\label{L-measure}
 Let $\{L_{tU}\}_{0\le t\le U}$ denote a multivariate LRB with marginal law $\nu$. Let the multivariate L\'evy process $\{L_t\}_{0\le t\le U}$, which generates the LRB, have density $\rho_t(x)$ for all $t\in(0,U]$. Under the measure $\mL$ defined by 
 \begin{equation}\label{l-mart}
  \ell_t^{-1}:=\frac{\rd\PR}{\rd\mL}\bigg\vert_{\F_t}=\int_{\R^m}\frac{\rho_{U-t}(z-L_{tU})}{\rho_U(z)}\,\nu(\rd z),
 \end{equation}
 the LRB $\{L_{tU}\}$ has the law of the generating L\'evy process for $t\in[0,U)$.
\end{proposition}

The verification of this proposition follows closely the results leading to Proposition 3.7 in Hoyle {\it et al}. \cite{hhm1}. The measure $\mL$ is rather useful for several calculations as we will see, shortly. LRBs, which have joint marginal law at $t=U$ and which are generated by $\mL$-independent L\'evy processes, are nevertheless independent under $\mL$. 

Next, we propose multi-factor pricing kernel models, and thus multi-factor asset price models, in the situation where the driving Markov process is a multivariate LRB. The construction of these models follows the technique presented at the end of Section \ref{Sec-II} and in Section \ref{Sec-cfmodels}.

\paragraph{\bf Exponential linear two-factor model with jumps.} We assume that the market filtration $\{\F_t\}$ is generated by a two-dimensional LRB, of which the first component is a Brownian random bridge,
\begin{equation*}
 L^{(1)}_{tU}=\sigma\,X^{(1)}_U\,t+\beta_{tU},
\end{equation*}
and the second component  is a gamma random bridge defined by 
\begin{equation*}
 L^{(2)}_{tU}=X^{(2)}_U\,\gamma_{tU}.
\end{equation*}
The Brownian bridge $\{\beta_{tU}\}$ and the gamma bridge $\{\gamma_{tU}\}$ are assumed $\mL$-independent of each other and also $\mL$-independent of the random variables $X^{(1)}_U$ and $X^{(2)}_U$. However, the two X random variables may be dependent and have {\it a priori} bivariate marginal law $\nu$. Next, we recall the bond pricing formula (\ref{M-bond}) where, this time,
\begin{eqnarray}\label{Y2fact}
 Y^{\mL}_{tT}&=&\int^{U-t}_{T-t}\int^{U-t}_{T-t}w(T,u_1-T+t,u_2-T+t)\nonumber\\ &\times&\E^{\mL}\left[F\left(t+u_1,t+u_2,L^{(1)}_{t+u_1,U},L^{(2)}_{t+u_2,U}\right)\,\big\vert\,L^{(1)}_{tU},L^{(2)}_{tU}\right]\rd u_1\rd u_2.
\end{eqnarray}
Here we replace $\M$ in (\ref{M-bond}) with $\mL$ to emphasise that the measure is changed to the $\mL$-measure. Since, for $t\in[0,U)$, the two LRB components each have the $\mL$-law of the corresponding underlying L\'evy process (Brownian motion and the gamma process, respectively), the conditional expectation simplifies considerably under $\mL$. That is,
\begin{align}
 &\E^{\mL}\left[F\left(t+u_1,t+u_2,L^{(1)}_{t+u_1,U},L^{(2)}_{t+u_2,U}\right)\,\big\vert\,L^{(1)}_{tU}=x_1,L^{(2)}_{tU}=x_2\right]\nonumber\\
 &=\E^{\mL}\left[F\left(t+u_1,t+u_2,\left(L^{(1)}_{t+u_1,U}-L^{(1)}_{tU}\right)+x_1,\left(L^{(2)}_{t+u_2,U}-L^{(2)}_{tU}\right)+x_2\right)\right].\nonumber
\end{align}
For the LRB components are $\mL$-independent, even though $X^{(1)}_U$ and $X^{(2)}_U$ may be assumed dependent, and for the $\mL$-laws of the LRB components are known, the probability densities of the increments in the above equation are also known. We thus have:
\begin{eqnarray}
 &&\mL\left[L^{(1)}_{t+u_1,U}-L^{(1)}_{tU}\in\rd y_1\right]=\frac{1}{\sqrt{2\pi\,u_1}}\exp\left(-\frac{y_1^2}{2\,u_1}\right)\rd y_1,\label{M1L1}\\
 &&\mL\left[L^{(2)}_{t+u_2,U}-L^{(2)}_{tU}\in\rd y_2\right]=\frac{\indi{1}\{y_2>0\}}{\Gamma[mu_2]}\,y_2^{mu_2-1}\exp(-y_2)\,\rd y_2,\label{M2L2}
\end{eqnarray}
where $m>0$ and $\Gamma[x]$ is the gamma function. In order to work out an explicit example, we need to specify $F(t,y_1,y_2)$ and the weight function $w(t,u_1,u_2)$. We choose the following:
\begin{align}\label{F2fact}
 &F\left(t+u_1,t+u_2,y_1+x_1,y_2+x_2\right)=\exp\left[a\left(y_1+x_1\right)-c\left(y_2+x_2\right)\right],\\
 &w(t,u_1,u_2)=\exp\left(-\tfrac{1}{2}a^2(t+u_1)\right)(c+1)^{m(t+u_2)},\label{w2fact}
\end{align}
where $a\in[-\infty,\infty)$, $c\ge 0$ are constants. We then insert (\ref{M1L1}) and (\ref{M2L2}), together with (\ref{F2fact}) and (\ref{w2fact}), in (\ref{Y2fact}) and calculate the integrals over $u_1$ and $u_2$. The result is:
\begin{align}\label{Y2fact-calc}
Y^{\mL}_{tT}=(U-T)^2(c+1)^{mt}\exp\left(a\,L^{(1)}_{tU}-c\,L^{(2)}_{tU}-\tfrac{1}{2}\,a^2 t\right).
\end{align}
The two-factor pricing kernel, jointly driven by a Brownian random bridge and a gamma random bridge, is thus given by a formula similar to (\ref{M-pk}) where the change-of-measure density martingale $\{M_t\}$, in the LRB context denoted $\{\ell_t\}$, is the reciprocal of (\ref{l-mart}) while $Y^{\mL}_{tt}$, $Y^{\mL}_{0t}$, and $Y^{\mL}_{00}$ are deduced from (\ref{Y2fact-calc}). As for the quadratic and exponential quadratic models analysed in Section \ref{Sec-cfmodels}, also this class of models for the bond price can be written in the form (\ref{Ab-bond}). We have:
\begin{equation}\label{2dim-bA-bond}
P_{tT}=\frac{P_{0T}+b(T)\,A^{\mL}_t}{P_{0t}+b(t)\,A^{\mL}_t}
\end{equation}
where, for $0\le t\le T<U$, 
\begin{align}
&P_{0t}=\frac{f_0(t)+(U-t)^2f_1(t)}{1+U^2f_1(0)},& &b(t)=\frac{(U-t)^2 f_1(t)}{1+U^2f_1(0)},&\nn\\
&A^{\mL}_t=(c+1)^{mt}\exp\left(a\,L^{(1)}_{tU}-c\,L^{(2)}_{tU}-\tfrac{1}{2}\,a^2 t\right)-1.&
\label{2dim-A}
\end{align}
One can show that $\{A^{\mL}_t\}$ is an $(\{\F_t\},\mL)$-martingale for $t\in[0,U)$. Then, the process $\{\ell_t A^{\mL}_t\}$ is an $(\{\F_t\},\PR)$-martingale, and the bond price process (\ref{2dim-bA-bond}) has a representation in terms of $(\{\F_t\},\PR)$-martingales, that is
\begin{equation}\label{2dim-bA-bond-P}
P_{tT}=\frac{P_{0T}\,\ell_t+b(T)\,A^{\PR}_t}{P_{0t}\,\ell_t+b(t)\,A^{\PR}_t},
\end{equation} 
where $A^{\PR}_t=\ell_t\,A^{\mL}_t$. Such models might be regarded as belonging to the finite-time equivalence class of Flesaker \& Hughston \cite{fh} bond price models. We note that pricing kernel models over infinite time, as in \cite{fh}, must be potentials of class D, see Rogers \cite{Ro}, Meyer \cite{mey}. In finite time, pricing kernel processes merely need to be positive $(\{\F_t\},\PR)$-supermartingales to ensure non-negative interest rates. Furthermore, the martingale processes underlying the price models arise endogenously from the pricing kernel structure (\ref{genpk}). Formula (\ref{genpk}) can be viewed as a ``machine'' that implicitly produces martingales for pricing formulae with a rational form. In Bermin \cite{ber}, the Flesaker-Hughston approach to bond pricing is revisited and it is shown how yield curves may be inverted for any short rate process consistent with bond price processes that have an exponentially-affine structure. 

In Section \ref{Sec-cfmodels}, explicit pricing models are derived, and one may ask at this point what the connection is between these pricing models and the ones specified in (\ref{2dim-bA-bond}). The link is a change of probability measure. Let us consider an $(\{\F_t\},\mL)$-martingale $\{A^{\mL}_t\}$ and an $(\{\F_t\},\M)$-martingale $\{A^{\M}_t\}$. Furthermore, we introduce an $(\{\F_t\},\mL)$-density-martingale $\{\eta_t\}_{0\le t<U}$ that changes the probability measure $\mL$ to the equivalent measure $\M$. We set $A^{\mL}_t=\eta_t\,A^{\M}_t$, and finally observe, for $0\le s\le t<U$, that
\begin{eqnarray*}
 \E^{\mL}\left[A^{\mL}_t\,\big\vert\,\F_s\right]=\E^{\mL}\left[\eta_t\,A^{\M}_t\,\big\vert\,\F_s\right]=\eta_s\,\E^{\M}\left[A^{\M}_t\,\big\vert\,\F_s\right]=\eta_s\,A^{\M}_s=A^{\mL}_s.
\end{eqnarray*}
This type of relation is also what connects (\ref{2dim-bA-bond}) and (\ref{2dim-bA-bond-P}).

\paragraph{\bf A useful formula.} The fact that an LRB has the law of its generating L\'evy process under the ``L\'evy probability measure'' $\mL$ can be exploited to derive asset price formulae expressed in terms of the characteristic function of a L\'evy process and a Fourier transform. We consider again the $\mL$-conditional expectation 
\begin{equation*}
 \E^{\mL}\left[F(t+u,L_{t+u,U})\,\vert\,L_{tU}\right],
\end{equation*}
where $\{L_{tU}\}$ is an LRB. We specify $F(t,x)$ by its Fourier transform $\widehat{F}(t,y)$, that is
\begin{equation*}
 F(t,x)=\int_{\R}\exp(-i\,xy)\widehat{F}(t,y)\,\rd y,
\end{equation*}
where $\widehat{F}(t,y)$ is selected such that $F(t,x)$ is positive and integrable. We then have:
\begin{equation*}
  \E^{\mL}\left[F(t+u,L_{t+u,U})\,\vert\,L_{tU}\right]= \E^{\mL}\left[\int_{\R}\exp(-iyL_{t+u,U})\,\widehat{F}(t+u,y)\,\rd y\,\bigg\vert\,L_{tU}\right].
\end{equation*}
Assuming that Fubini's Theorem is herewith satisfied, we swap the expectation with the integral, and obtain
\begin{equation*}
 \E^{\mL}\left[F(t+u,L_{t+u,U})\,\vert\,L_{tU}\right]=\int_{\R}\E^{\mL}\left[\exp\left(-iyL_{t+u,U}\right)\,\big\vert\,L_{tU}\right]\widehat{F}(t+u,y)\,\rd y.
\end{equation*}
Since, under $\mL$, the LRB has the law of the underlying L\'evy process for $t\in[0,U)$, we can calculate the conditional expectation by recalling that the increments of a L\'evy process are independent and stationary. The result is:
\begin{equation*}
 \E^{\mL}\left[\exp\left(-iyL_{t+u,U}\right)\,\big\vert\,L_{tU}\right]=\exp\left(-iyL_{tU}\right)\E^{\mL}\left[\exp\left(-iyL_{uU}\right)\right].
\end{equation*}
The expectation on the right-hand-side of the equation above is the generating function of a L\'evy process. We denote the characteristic function of a L\'evy process by $\Psi(y)$, and thus write
\begin{equation*}
 \E^{\mL}\left[\exp\left(-iyL_{uU}\right)\right]=\exp\left[-u\,\Psi(y)\right],
\end{equation*}
for $u\in[0,U)$. This leads to
\begin{equation*}
 \E^{\mL}\left[F(t+u,L_{t+u,U})\,\vert\,L_{tU}\right]=\int_{\R}\exp\left[-iyL_{tU}-u\,\Psi(u)\right]\widehat{F}(t+u,y)\,\rd y,
\end{equation*}
and hence to the useful formula
\begin{align}\label{useful-formula}
 Y^{\mL}_{tT}&=\int^{U-t}_{T-t}w(T,u-T+t)\,\E^{\mL}\left[F(t+u,L_{t+u,U})\,\vert\,X_{tU}\right]\rd u,\nonumber\\
	    &=\int^{U-t}_{T-t}w(T,u-T+t)\int_{\R}\exp\left[-iyL_{tU}-u\,\Psi(u)\right]\widehat{F}(t+u,y)\,\rd y\,\rd u.
\end{align}
Expression (\ref{useful-formula}) is valid also in the multi-factor case, in which the LRB $\{L_{tU}\}$ is a multi-dimensional vector. The elements of the LRB vector may be dependent through their terminal marginal laws as considered at the beginning of this section. The model (\ref{Y2fact-calc}) may be derived as a special case of the formula (\ref{useful-formula}). In order to obtain bond price models, (\ref{useful-formula}) is inserted in (\ref{M-bond}) to replace $Y^{\M}_{tT}$.

We conclude this section by producing multi-dimensional and multi-factor asset price processes. For the rest of this section, the following process $\{Y^{(i)}_{tT}\}$ is defined under the $\mL$-measure. For $i=1,2,\ldots,n$, let 
\begin{equation*}
 Y^{(i)}_{tT}=\int^{U-t}_{T-t}\E^{\mL}\left[F_i(t+u,L_{t+u,U})\,\vert\,L_{tU}\right]w_i(T,u-T+t)\,\rd u,
\end{equation*} 
where $F_i(t,x)$ is a positive and integrable function, and $w_i(t,u)$ is a weight function such that the combination of the two ensure $Y^{(i)}_{tT}<\infty$ for all $t$. We emphasise that the Markov process $\{L_{tU}\}_{0\le t\le U}$ may be multi-dimensional. Then, the following is a multi-dimensional and multi-factor model for the bond price:
\begin{equation}\label{bond-Ysum}
 P_{tT}=\frac{P_{0T}+\sum^n_{i=1}y_i(T)\left(Y^{(i)}_{tT}-Y^{(i)}_{0T}\right)}{P_{0t}+\sum^n_{i=1}y_i(t)\left(Y^{(i)}_{tt}-Y^{(i)}_{0t}\right)},
\end{equation}
where
\begin{equation*}
 y_i(t)=\frac{f_i(t)}{1+\sum_{i=1}^n Y^{(i)}_{00}}. 
 \end{equation*}
These models can be extended further. Let us assume that the multivariate LRBs generating the market filtration $\{\F_t\}$ are driven by $\mL$-independent L\'evy processes and may have joint terminal marginal distribution. Then, the product of an $(\{\F_t\},\mL)$-supermartingale is again an $(\{\F_t\},\mL)$-supermartingale. This leads us to the construction of higher-order asset pricing models, hereunder applied to the pricing of bonds. We generalise the bond price model (\ref{2dim-bA-bond}): For $0\le t\le T$ and $N\in\mathbb{N}$,
\begin{equation}\label{superbond}
 P_{tT}=\frac{P_{0T}+\sum^N_{i=1}\Lambda^{(i)}_{tT}}{P_{0t}+\sum^N_{i=1}\Lambda^{(i)}_{tt}},
\end{equation}
where $\Lambda^{(i)}_{tT}=$
\begin{align}\label{super-sum}
 &\sum^{n_i-m_i}_{j_{n_i-m_i}=1}b^{(i)}_{j_{n_i-m_i}}(T)\,A_t^{(i,\,j_{n_i-m_i})}\hspace{-.25cm}\sum^{n_i-(m_i-1)}_{j_{n_i-(m_i-1)}=j_{n_i-m_i}+1}b^{(i)}_{j_{n_i-(m_i-1)}}(T)\,A_t^{(i,\,j_{n_i-(m_i-1)})}\nn\\ 
 &\cdots\ \sum^{n_i-1}_{j_{n_i-1}=j_{n_i-2}+1}b^{(i)}_{j_{n_i-1}}(T)\,A_t^{(i,\,j_{n_i-1})}\sum^n_{j_{n_i}=j_{{n_i}-1}+1}b^{(i)}_{j_{n_i}}(T)\,A_t^{(i,\,j_{n_i})}
\end{align}
for $n_i\ge m_i\in\mathbb{N}$. By setting $T=t$ in (\ref{super-sum}) one obtains $\{\Lambda_{tt}\}$. The deterministic functions $b^{(i)}_1(t),b^{(i)}_2(t),\ldots,b^{(i)}_n(t)$ are specified such that the pricing kernel underlying the price model is a positive supermartingale. The processes $\{A^{(i,j)}_t\}_{0\le t<U}$ and $\{A^{(i,k)}_t\}_{0\le t<U}$ are $(\{\F_t\},\mL)$-martingales, and these are $\mL$-independent for $j\neq k$. For instance, for $N=1$, $n_1=3$ and $m_1=2$, one obtains the third-order model
\begin{equation}\label{3rd-order}
 P_{tT}=\frac{P_{0T}+b_{123}(T)\,A_t^{(1)}\,A_t^{(2)}\,A_t^{(3)}}{P_{0t}+b_{123}(t)\,A_t^{(1)}\,A_t^{(2)}\,A_t^{(3)}},
\end{equation}
where $b_{123}(t)=b_1(t)\,b_2(t)\,b_3(t)$ for $0\le t\le T$. For $N=1$, $n_1=3$ and $m_1=1$, we have the second-order model
\begin{equation}\label{2nd-order}
 P_{tT}=\frac{P_{0T}+b_{12}(T)\,A^{(1)}_t\,A^{(2)}_t+b_{13}(T)\,A^{(1)}_t\,A^{(3)}_t+b_{23}(T)\,A^{(2)}_t\,A^{(3)}_t}{P_{0t}+b_{12}(t)\,A^{(1)}_t\,A^{(2)}_t+b_{13}(t)\,A^{(1)}_t\,A^{(3)}_t+b_{23}(t)\,A^{(2)}_t\,A^{(3)}_t},
\end{equation}
where $b_{ij}(t)=b_j(t)b_k(t)$ for $j\neq k$ and $0\le t\le T$. In order to lighten the notation, the $i$-index is suppressed in (\ref{3rd-order}) and (\ref{2nd-order}) since we have only one type of higher-order term in the sum over $i$. For $N=2$, $n_1=3$ and $m_1=2$, $n_2=3$ and $m_2=1$, a combination of third-order and second-order summands drives the bond price. We note that the construction of higher-order pricing formulae is not limited to models driven by LRBs. The pricing kernel model (\ref{genpk}) can be used to construct higher-order price models driven by other Markov processes. Higher-order models gain in importance when considering general asset pricing including dependences across several types of asset. For instance, models such as (\ref{3rd-order}) and (\ref{2nd-order}) might be applied in the pricing of bond portfolios where the portfolio assets share one or more ``drivers'' in common, be them the $A$-processes of various types (e. g. quadratic, exponential linear and quadratic, etc.) and/or the underlying Markov processes, perhaps also of various types of probability laws. Another situation in which one can foresee higher-order models to be useful is when a financial instrument is exposed to various sectors of a financial market. We can think of a portfolio of shares, of which prices are discounted and thus certainly linked to the bond market. Then there are dependences across the shares composing the portfolio; perhaps one is also interested to hedge the portfolio with positions in a shares index, and so on. These are all situations where the degrees of freedom of the higher-order models are expected to be useful, in particular for the modelling of portfolio dependences. The next section is devoted to the pricing of general assets, and we shall keep in mind that higher-order models can be constructed also for general financial instruments.
\section{General asset pricing in finite time}\label{Gen-asset-p}
The pricing formula (\ref{pf}) implies that the price process of an asset has the martingale property with respect to the market filtration $\{\F_t\}$ and the real probability measure $\PR$. There are several ways to construct $(\{\F_t\},\PR)$-martingales; we consider however a natural method within the framework developed thus far. For some fixed $T$, we denote by $\{m_{tT}\}_{0\le t\le T}$ an $(\{\F_t\},\PR)$-martingale, and write 
\begin{equation}\label{pf-martingale}
 \pi_t S_{tT}=m_{tT}.
\end{equation}
\begin{definition}\label{ZtT}
 Let $\{Z_{tT}\}_{0\le t\le T<U}$ be defined by
\begin{equation}\label{Zproc}
 Z_{tT}=\int^{U-t}_{T-t}\E^{\PR}\left[G\left(t+u,X_{t+u}\right)\,\vert\,X_t\right]\psi(T,u-T+t)\rd u,
\end{equation}
where the deterministic function $G(t,x)$ is measurable, $\psi(t,u)$ is a deterministic and measurable function with the property $\psi(t,u-s)=\psi(t-s,u)$ for $s\le t\wedge u$. The combination $G(t,x)$ and $\psi(t,u)$ is such that it ensures $Z_{tT}<\infty$ for all $t$.
\end{definition}
\begin{proposition}
Let $0\le t\le T<U$, and let $g_0(T)$ and $g_1(T)$ be deterministic functions. Then, for each fixed $T$, 
\begin{equation}\label{m-mart}
 m_{tT}=g_0(T)+g_1(T)Z_{tT}
\end{equation}
is an $(\{\F_t\},\PR)$-martingale.
\end{proposition}
Proof. This proposition can be proven by following the steps in the proof of Proposition 2.2 in Akahori \& Macrina \cite{am}. We observe that $\psi(t,u-s)=\psi(t-s,u)$ implies $\psi(t,u)=\psi(t+u)$. \hfill$\Box$
\\

In contrast to $F(t,x)$ and $w(t,u)$ in (\ref{genpk}), we require that neither $G(t,x)$ nor $\psi(t,u)$ be positive functions. This is to include the pricing of assets that are not of limited liability and thus do not necessarily have positive prices. We now have the necessary ingredients in order to propose the following class of asset price models.
\begin{lemma}\label{StT}
 Let $\{\pi_t\}$ and $\{m_{tT}\}$ be the processes (\ref{pricingkernel}) and (\ref{m-mart}), respectively. Then the asset price model (\ref{pf-martingale}) takes the form
\begin{equation}\label{StTwith Z}
 S_{tT}=\frac{S_{0T}+z(T)\left(Z_{tT}-Z_{0T}\right)}{P_{0t}+y(t)\left(Y_{tt}-Y_{0t}\right)},
\end{equation}
where $z(T)=g_1(T)/\pi_0$.
\end{lemma}
Proof. The relation (\ref{StTwith Z}) follows from (\ref{pf-martingale}) by inserting (\ref{pricingkernel}) and (\ref{m-mart}). The degree of freedom $g_0(T)$ can be calibrated to the asset price $S_{0T}$ at time 0 via
\begin{equation*}
 g_0(T)=S_{0T}\pi_0-g_1(T)Z_{0T},
\end{equation*}
where $\pi_0=1+f_1(0)Y_{00}$. \hfill$\Box$
\\

Since the martingale family $\{m_{tT}\}$ is not necessarily positive-valued, the price process $\{S_{tT}\}$ is neither. We deliberately keep this level of generality, as opposed to requiring assets to have positive prices, since we might wish to consider also the pricing of portfolios of which value might become negative at times. As at the beginning of Section \ref{Sec-cfmodels}, we next derive general asset price models for which semi-explicit (possibly up to numerical root-finding) expressions for derivatives can be computed. Depending on the form of the functions $Y_{tt}$ and $Z_{tT}$, the ``useful formula" (\ref{useful-formula}) can be applied to compute option prices. We also apply the results in Proposition \ref{Prop-M-pk}.
\begin{lemma}\label{Ab-S}
Given the class of asset price models (\ref{StTwith Z}), there exist equivalent measures $\M$, induced by a change-of-measure density $(\{\F_t\},\PR)$-martingale $\{M_t\}_{0\le t<U}$, $(\{\F_t\},\M)$-martingales $\{A^{(i)}_t\}^{i=1,2}_{0\le t<U}$ with $A^{(i)}_0=0$, and deterministic functions $b_i(t)$ such that the asset price process $\{S_{tT}\}_{0\le t\le T<U}$ takes the form
\begin{equation}\label{equitypp}
 S_{tT}=\frac{S_{0T}+b_1(T)\,A^{(1)}_t}{P_{0t}+b_2(t)\,A^{(2)}_t}.
\end{equation}
The asset price $S_{tT}$ is the value at time $t$ of the cash flow 
\begin{equation}
 S_{TT}=\frac{S_{0T}+b_1(T)\,A^{(1)}_T}{P_{0T}+b_2(T)\,A^{(2)}_T} 
\end{equation}
at time $T$, where $\{S_{tT}\}$ is quoted in units of the pricing kernel process
\begin{equation}
\pi_t=\frac{\pi_0}{M_0}\left[P_{0t}+b_2(t)\,A^{(2)}_t\right]M_t.
\end{equation} 
\end{lemma}
Proof. The expression (\ref{equitypp}) is obtained by computing the expectation in 
\begin{equation*}
S_{tT}=\frac{1}{\pi_t}\,\E^{\PR}\left[\pi_T\,S_{TT}\,\vert\,\F_t\right]  
\end{equation*}
where, by use of $\{M_t\}$, the measure $\PR$ is first changed to $\M$ in order to exploit the martingale property of $\{A^{(1)}_t\}_{0\le t<U}$ under $\M$. A first example proving existence follows immediately. A second example is given at the end of this section.\hfill$\Box$

\paragraph{\bf Asset price diffusion with stochastic discounting.} Let the market filtration $\{\F_t\}$ be generated by two Brownian random bridges $\{L^{(i)}_{tU}\}$, $i=1,2$, which are constructed in terms of possibly dependent random variables $X^{(i)}_{U}$. We consider the quadratic model (\ref{Quad-Ab}) for the pricing kernel $\{\pi_t\}$ modelled in terms of $\{A^{(2)}_t\}$ and driven by $\{L^{(2)}_{tU}\}$, and the exponential quadratic model (\ref{ExpQuad-Ab}) for the component modelled in terms of $\{A^{(1)}_t\}$ and driven by $\{L^{(1)}_{tU}\}$. In such a case, one obtains an asset price process of the form
\begin{equation}\label{S-example}
 S_{tT}=\frac{S_{0T}+\frac{\frac{1}{\eta}(U-T)^{\eta}U^{1/2}g_1(T)}{4U\left[1+(1/12)\,f_1(0)U^3\right]}\left[\sqrt{1-t/U}\,\exp\left(\frac{L^{(1)\ 2}_{tU}}{2(U-t)}\right)-1\right]}{P_{0t}+\frac{(U-t)^4 f_1(t)}{4U\left[1+(1/12)\,f_1(0)U^3\right]}\left[\frac{U}{(U-t)^2}\,L^{(2)\ 2}_{tU}-\frac{t}{U-t}\right]},
\end{equation}
where $P_{0t}$ is specified in (\ref{Quad-Ab}) and
\begin{equation}\label{S0-example}
S_{0T}=\frac{g_0(T)+\frac{1}{\eta}(U-T)^{\eta}U^{1/2}g_1(T)}{1+\frac{1}{12}U^3 f_1(0)}.
\end{equation}
The price process (\ref{S-example}) can be generalised to an incomplete market setup by considering a vector Brownian random bridge $\{\bar{L}^{(1)}_{tU},\bar{L}^{(2)}_{tU},\ldots,\bar{L}^{(n)}_{tU}\}$. Then, for instance, we might set $L^{(1)}_{tU}=\{\bar{L}^{(1)}_{tU},\bar{L}^{(2)}_{tU},\ldots,\bar{L}^{(m)}_{tU}\}$ and $L^{(2)}_{tU}=\{\bar{L}^{(1)}_{tU},\bar{L}^{(2)}_{tU},\ldots,\bar{L}^{(n)}_{tU}\}$, $m\le n$. Another way is proposed in (\ref{superbond}). 

\begin{figure}[H]
\begin{center}
\includegraphics[scale=.5, angle=0]{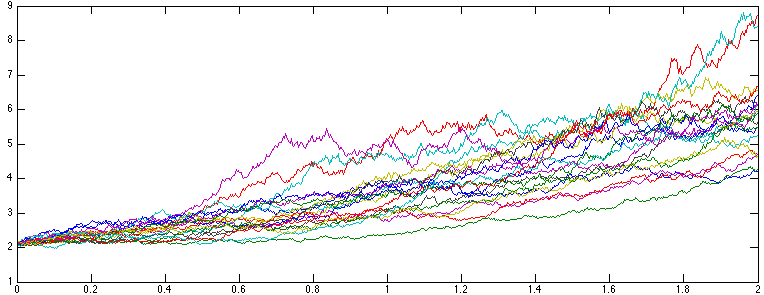}
\caption{Simulation of the price process (\ref{S-example}). The deterministic functions $f_0(t)$ and $f_1(t)$ are defined as in (\ref{f0f1}), whereas $g_0(t)$ and $g_1(t)$ decay exponentially with a constant damping rate. $\eta=1$. Further details on how such processes are simulated are provided in the paragraph preceding Figure \ref{figGBc1}.}
\vspace{-.25cm}
\label{fig-AP-EQQ}
\end{center}
\end{figure}

\paragraph{\bf Dynamical equations.} We consider price processes of the form (\ref{equitypp}) which are adapted to a filtration $\{\F_t\}$ generated by Brownian random bridges
\begin{equation}
L^{(i)}_{tU}=\sigma_i\,X^{(i)}_U\,t+\beta^{(i)}_{tU}.
\end{equation}
While $X^{(i)}_U$ and $\{\beta^{(i)}_{tU}\}$ are assumed independent and $\sigma_i$ is constant, we may take the Brownian bridges $\{\beta^{(i)}_{tU}\}$ to be correlated, here. We further assume that the processes $\{A^{(i)}_t\}_{0\le t<U}$ satisfy
 \begin{align}\label{Ito-A-S}
  &\rd A^{(i)}_t=\nu^{(i)}_t\left(\rd W^{\PR\ (i)}_t+\vartheta^{(i)}_t\rd t\right)\nn\\
  &A^{(i)}_0=0,
 \end{align}
 where $\{\nu^{(i)}_t\}$ is $\{\F_t\}$-adapted and
\begin{eqnarray}\label{WP-inno}
  &&\vartheta^{(i)}_t=\frac{\sigma_i\,U}{U-t}\,\E^{\PR}\left[X^{(i)}_U\,\big\vert\,\F_t\right],\nn\\
  &&\rd W^{\PR\ (i)}_t=\rd L^{(i)}_{tU}-\frac{1}{U-t}\left(\sigma_i\,U\,\E^{\PR}\left[X^{(i)}_U\,\big\vert\,\F_t\right]-L^{(i)}_{tU}\right)\rd t.
 \end{eqnarray}
It then follows, by the Ito Formula, that, on $0\le t\le T<U$, the price process (\ref{equitypp}) satisfies the SDE
 \begin{equation}\label{S-dyn}
  \frac{\rd S_{tT}}{S_{tT}}=\left(r_t+\lambda_t\Sigma_{tT}\right)\rd t+\Sigma_{tT}\,\rd W^{\PR}_t,
 \end{equation}
where 
\begin{align}
 &r_t=-\,\frac{\partial_t P_{0t}+A^{(2)}_t\,\partial_t{b}_2(t)}{P_{0t}+b_2(t)\,A^{(2)}_t},&
 &\lambda_t=\begin{pmatrix}
              \vartheta_t^{(1)}-\rho_{ij}\,\nu_t^{(2)}\,\frac{b_2(t)}{P_{0t}+b_2(t)\,A^{(2)}_t}\\
              \\
              \vartheta_t^{(2)}-\nu_t^{(2)}\,\frac{b_2(t)}{P_{0t}+b_2(t)\,A^{(2)}_t}
             \end{pmatrix},&\nn\\
 &\Sigma_{tT}=\begin{pmatrix}
              \frac{b_1(T)\nu^{(1)}_t}{S_{0T}+b_1(T)\,A^{(1)}_t}\\
              \\
              -\,\frac{b_2(t)\nu^{(2)}_t}{P_{0t}+b_2(t)\,A^{(2)}_t}
              \end{pmatrix}.& \nn           
\end{align} 
The process $W^{\PR}_t=(W^{\PR\ (1)}_t,W^{\PR\ (2)}_t)$ is a two-dimensional $(\{\F_t\},\PR)$-Brownian motion where $\rd W^{\PR\ (i)}_t\rd W^{\PR\ (j)}_t=\rho_{ij}\,\rd t$ for $i\neq j$, $\rho_{ij}\in[-1,1)$, and $\rd W^{\PR\ (i)}_t\rd W^{\PR\ (j)}_t=\rd t$ for $i=j$. 
\begin{remark}
The example (\ref{S-example}) satisfies (\ref{S-dyn}) where $A^{(1)}_t$, $b_1(t)$ and $S_{0T}$ are respectively specified in $(\ref{ExpQuad-Ab})$ and $(\ref{S0-example})$, and $A^{(2)}_t$, $b_2(t)$ and $P_{0t}$ are determined in (\ref{Quad-Ab}). Furthermore,
\begin{align*}
&\nu^{(1)}_t=\frac{\left(\frac{U-t}{U}\right)^{1/2}}{U-t}\,L^{(1)}_{tU}\,\exp\left[\frac{L^{(1)\ 2}_{tU}}{2(U-t)}\right],& &\nu^{(2)}_t=\frac{2U}{(U-t)^2}\,L^{(2)}_{tU}.&
\end{align*}   
Another example is, of course, where $\{A^{(1)}_t\}$ and $\{A^{(2)}_t\}$ respectively belong to the quadratic and exponential quadratic class, instead. Generalised versions of (\ref{S-dyn}) can be obtained in which the driving random Brownian bridges are multidimensional and the price process (\ref{equitypp}) is constructed by higher-dimensional systems as in (\ref{3rd-order}) or (\ref{2nd-order}). We imagine multidimensional examples being useful when a more refined dependence structure between the ``equity component" and the discount factor is necessary, and in particular when modelling portfolio assets. Also, the market filtration might be modelled by different random Gaussian bridges which are likely to suggest different classes of $\{A_t\}$-processes.
\end{remark}
\begin{remark}
The processes $\{\nu^{(i)}_t\}$, $i=1,2$, are determined by the specific choice of the models at the basis of the processes $\{A^{(i)}_t\}$, and thus are model-specific. Inside the market price of risk vector $\{\lambda_t\}$, one notices the correlation parameter $\rho_{ij}$ that arises from the $\rd A^{(i)}_t\rd A^{(j)}_t$-term containing $\rd W^{\PR\ (i)}_t\rd W^{\PR\ (j)}_t$ that, following (\ref{WP-inno}), yields $\rd L^{(i)}_{tU}\,\rd L^{(j)}_{tU}=\rd\beta^{(i)}_{tU}\,\rd\beta^{(j)}_{tU}=\rho_{ij}\rd t$. 
\end{remark}
\begin{remark}
The dynamics (\ref{S-dyn}) can be transformed to obtain the risk-neutral dynamical equation of the asset price process $\{S_{tT}\}_{0\le t\le T}$. We have:
\begin{equation}\label{S-Qdyn}
 \frac{\rd S_{tT}}{S_{tT}}=r_t\rd t+\Sigma_{tT}\,\rd W^{\Q}_t,
\end{equation}
where $\rd W^{\Q}_t=\rd W^{\PR}_t+\lambda_t\rd t$ is the risk-neutral Brownian motion defined in terms of the $\PR$-Brownian motion $\{W^{\PR}_t\}$ and the market price of risk process $\{\lambda_t\}$. The solution to the stochastic differential equation (\ref{S-Qdyn}) has the familiar $\Q$-log-normal form 
\begin{equation}\label{log-normal-mod}
 S_{tT}=S_{0T}\exp\left(\int^t_0\left(r_s-\tfrac{1}{2}\,\Sigma_{sT}^2\right)\rd s+\int^t_0\Sigma_{sT}\,\rd W^{\Q}_s\right).
\end{equation}
How one may obtain ``finite-time Black-Scholes-type models'' from (\ref{log-normal-mod}) can be deduced by consulting Brody {\it et al}. \cite{bhm}, Section 9. In addition, appropriate choices for the functions $F(x)$, $G(x)$, and the related weight functions will need to be made.
\end{remark}
\begin{remark}
The solution to the stochastic differential equation (\ref{S-dyn}) is a positive-valued process, namely (\ref{log-normal-mod}) as written in its risk-neutral form. Recalling Lemma (\ref{StT}), or in particular Lemma (\ref{Ab-S}), one may wonder what ensures the positivity of the constructed asset price process. The answer is: no such condition is imposed, in general. However, the diffusion price processes (\ref{S-dyn}) is based on the Ito-dynamics (\ref{Ito-A-S}), which in combination with the rational form of $\{S_t\}$, results in a positive price process. It is perhaps surprising that the SDE (\ref{S-dyn}) is solved by an explicit expression of the form (\ref{equitypp}), for instance by (\ref{S-example}), which might be hard to guess in its form (\ref{log-normal-mod}) or in the equivalent one under $\PR$.
\end{remark}
\paragraph{\bf Asset price dynamics with heavy tails and stochastic discounting.} We now give one more example of an asset price process (\ref{equitypp}), which is driven by three different types of L\'evy random bridges. Let the market filtration be generated by (i) a stable-1/2 random bridge $\{L^{(1)}_{tU}\}$, (ii) two gamma random bridges $\{L^{(2)}_{tU}\}$ and $\{L^{(3)}_{tU}\}$, and (iii) a Brownian random bridge $\{L^{(4)}_{tU}\}$. For details about 1/2-stable random bridges, we refer to Hoyle {\it et al}. \cite{hhm2}. The four random bridges are assumed to be independent under the ``L\'evy probability measure'' $\mL$, which is not to say that the marginals at time $t=U$ of the four LRBs are independent (as explained below Proposition \ref{L-measure}). Next we make use of Definition \ref{ZtT} and Proposition \ref{StT}. It is convenient to take the expectations under an auxiliary probability measure as presented in Proposition \ref{Prop-M-pk}. Given that in this example the market filtration is generated by four LRBs, we choose to compute the expectations under the ``L\'evy probability measure'' $\mL$, under which the LRBs have the law of the generating underlying L\'evy processes. For the function $G(t+u,X_{t+u})$ in Definition (\ref{ZtT}), we set
 \begin{equation}\label{G-1/2}
  G\left(t+u_1,t+u_2,L^{(1)}_{t+u_1,U},L^{(2)}_{t+u_2,U}\right)=\exp\left(-\kappa L^{(1)}_{t+u_1,U}+cL^{(2)}_{t+u_2,U} \right)
 \end{equation}
 where $\kappa\ge 0$, $0\le c\le 1$. The Laplace transform of the stable-1/2 subordinator $\{L_t\}$, which is a L\'evy process, is 
\begin{equation*}
 \E\left[\exp\left(-\kappa L_t\right)\right]=\exp\left(-\frac{\alpha\sqrt{\kappa}}{\sqrt{2}}\,t\right)
\end{equation*}
where $\alpha>0$ is the activity parameter that features in the subordinator's density function
\begin{equation}\label{1/2density}
\rho_t(y)=\indi{1}\{y>0\}\frac{\alpha\,t}{\sqrt{2\pi}\,y^{3/2}}\,\exp\left(-\frac{\alpha^2\,t^2}{2\,y}\right).
\end{equation}
We now calculate (\ref{Zproc}) with the specification (\ref{G-1/2}) and by setting
\begin{equation*}
 \psi(t+u)=\widetilde{\psi}(t,u_1,u_2)=\exp\left(\frac{\alpha\sqrt{\kappa}}{\sqrt{2}}(t+u_1)\right)\left(1-c\right)^{m(t+u_2)}
\end{equation*}
where $m>0$. The result is:
\begin{equation*}
 Z_{tT}=(U-T)^2\exp\left(-\kappa L^{(1)}_{tU}+\frac{\alpha\sqrt{\kappa}}{\sqrt{2}}\,t\right)(1-c)^{mt}\exp\left(cL^{(2)}_{tU}\right).
\end{equation*}
In this case, the price process $\{S_{tT}\}$ given in (\ref{StTwith Z}) can be written in the form (\ref{equitypp}). For the denominator, we choose a discount factor of the kind (\ref{2dim-bA-bond}). We obtain the following price process for, e. g., equity:
\begin{equation}\label{jump-S}
S_{tT}=\frac{S_{0T}+b_1(T)\,A^{(1)}_t}{P_{0t}+b_2(t)\,A^{(2)}_t}, 
\end{equation}
where, for $\eta\ge 0$, $a\in[-\infty,\infty)$, $q>0$, and $0\le t\le T<U$, we have
\begin{align}
&S_{0T}=\frac{g_0(T)+(U-T)^2g_1(T)}{1+U^2 f_1(0)},& &b_1(T)=\frac{(U-T)^2 g_1(T)}{1+U^2 f_1(0)},&\nn\\
&A^{(1)}_t=(1-c)^{mt}\exp\left(-\kappa L^{(1)}_{tU}+\frac{\alpha\sqrt{\kappa}}{\sqrt{2}}\,t+cL^{(2)}_{tU}\right)-1,&\nn\\
\label{jump-A1}\\
&P_{0t}=\frac{f_0(t)+(U-t)^2f_1(t)}{1+U^2 f_1(0)},& &b_2(t)=\frac{(U-t)^2 f_1(t)}{1+U^2 f_1(0)},&\nn\\ 
&A^{(2)}_t=(\eta+1)^{qt}\exp\left(-\eta L^{(3)}_{tU}+aL^{(4)}_{tU}-\frac{1}{2}a^2 t\right)-1.&\label{jump-A2}
\end{align}

\begin{figure}[H]
\begin{center}
\includegraphics[scale=.365, angle=0]{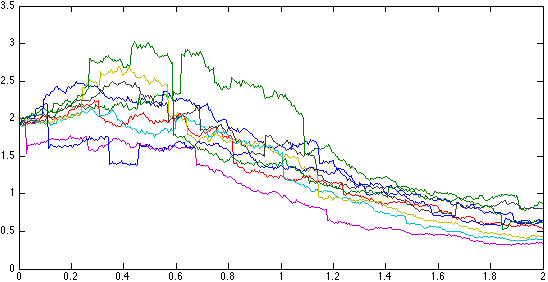}
\includegraphics[scale=.365, angle=0]{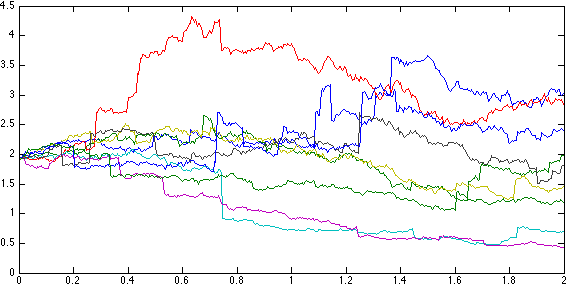}
\caption{Two distinct simulations on the price process (\ref{jump-S}) specified by (\ref{jump-A1}) and (\ref{jump-A2}). The four LRBs have four possible outcomes at time $U$, and their terminal marginal distribution is assumed to be joint. The stable-1/2 LRB is simulated as in Hoyle {\it et al.} \cite{hhm2}. The deterministic functions are the same as in Figure \ref{fig-AP-EQQ}. Further details on how such processes are simulated are provided in the paragraph preceding Figure \ref{figGBc1}.}
\vspace{-.25cm}
\label{fig-AP-EQQ}
\end{center}
\end{figure}

Another example could be constructed by generating asset price models driven by ``VG random bridges'' and quoted in units of the natural numeraire at the basis of the model (\ref{2dim-bA-bond}). In the case that the economic factors modelled by the random variables $L^{(i)}_{UU}$ are dependent, one obtains simple dependence structures between the dynamics of the equity and the associated discount bond system that determines the discount rate in the financial market. The more advanced dependence models introduced in the next section can also be applied to model interactions between different segments of a financial market.    
\section{Spiralling debt and its impact on international bond markets}\label{defic-mod}
We now address in more detail the pricing of sovereign bonds. Even though the majority of sovereign bonds pay coupons, we focus on discount bonds, for convenience. This simplification does not affect the view taken or the problem we intend to tackle here. The emphasis is shifted on the value of a sovereign bond that should reflect the level of economic health of the issuing country. News regarding the bond market over the last few years has constantly reminded us that investors frequently balance the capability of a sovereign economy to grow vis-a-vis the amount of accumulated debt held at any one time. We choose this point of view, and wonder how to construct asset pricing models, which take into account at least some of this perspective. 

We consider a simple model for the economic structure of a country. We assume that a central government has a source of income, for instance taxes and the revenues of state-owned companies. On the other hand it also has expenditures in order, for instance, to finance armed forces, public education, a public health system, and other welfare. While the difference between income and expenditures fluctuates over the course of time, we assume that it is unlikely, at least in a well-run and periodically well-assessed economy, that this difference spikes for the better or for the worse. If it were the case, then we might see an economy's growth rate move from 1\% to 10\% within a few months, or a drop in the growth rate by a similar amount in the same time span. It is more likely though that a central government has to step-in to cover huge unexpected losses due to, e. g., the unfolding of an international financial crisis, domestic or international wars, natural catastrophes, and other calamities hard to predict 
and with disastrous impact on the economic health of a country. So, in addition to the ``structural'' income and outcomes of an economy, we consider the accumulation of significant debt due to severe losses, which may very well make the level of financial stress of an economy ``jump''. We model the structural part of the various cash flows of an economy by a Brownian random bridge
\begin{equation*}
 L^{(1)}_{tU}=\sigma t X_U+\beta_{tU},
\end{equation*}
where $X_U$ may represent the economic wealth of a country at a future time $U$. We model the spiralling cumulative debt amassed by a sovereign country in the time interval $[0,U]$ by a gamma random bridge $\{L^{(2)}_{tU}\}$. The random total (extraordinary) debt amassed by time $U$ is modelled by $L^{(2)}_{UU}=X^{(2)}_U$. Its distribution can be arbitrarily specified. We imagine that the ``structural'' or ``non-crisis'' balance $L^{(1)}_{UU}$ is dependent on the total debt (losses) $L^{(2)}_{UU}$ accumulated by time $U$. For instance, a sovereign government may decide at time U to make substantial cuts to the expenditures for public welfare if the amount of ``extraordinary losses (debt)'' will have spiralled by time $U$ beyond what is perceived to be manageable. Therefore $L^{(1)}_{UU}$ and $L^{(2)}_{UU}$ are assumed to have a joint marginal distribution, and this means that we are in the same modelling environment as in Section \ref{Sec-incompmarket}.

The bond pricing model presented next is one of the simplest, though still rich enough to capture the desiderata within this discussion. One can of course choose to develop more sophisticated models. We choose a class of bond price models similar to (\ref{2dim-bA-bond}), and follow the steps from (\ref{Y2fact}) to (\ref{2dim-A}) with one minor change in equation (\ref{F2fact}). We consider
\begin{align}
 &F(t+u_1,t+u_2,y_1+x_1,y_2+x_2)=\exp\left(-a(y_1+x_1)+c(y_2+x_2)\right),& \nn\\
 &w(t,u_1,u_2)=\exp\left(-\frac{a^2}{2}(t+u_1)\right)(1-c)^{m(t+u_2)}.\nn
\end{align}
The reason for the change in the sign is that this way the losses will be recognised as downward jumps in the time series of the bond price. We emphasise that expectations are computed under the $\mL$-measure, under which LRBs with joint terminal marginals are nevertheless independent and inherit the law of the generating L\'evy processes. The bond price is then given by
\begin{equation}\label{debt-bond}
 P_{tT}=\frac{P_{0T}+b(T)\,A^{\mL}_t}{P_{0t}+b(t)\,A^{\mL}_t},
\end{equation}
where, for $a\ge 0$, $0\le c\le 1$, $m>0$, we have
\begin{align}
 &P_{0t}=\frac{f_0(t)+(U-t)^2f_1(t)}{1+U^2f_1(0)},& &b(t)=\frac{(U-t)^2 f_1(t)}{1+f_1(0)U^2},&\nn\\ 
 &A^{\mL}_t=(1-c)^{mt}\exp\left(-a\,L^{(1)}_{tU}-\tfrac{1}{2}a^2 t+c\,L^{(2)}_{tU}\right)-1,& &0\le t\le T<U.&
\end{align}

\paragraph{\bf Dependence in international markets.} The effects of losses getting out of control are not confined to one's domestic economy. Especially in a global financial market, the deterioration of an economy's health exposes, for instance, foreign creditors holding debt of the distressed economy to higher credit risk. Bond markets are global ``debt networks'' linking several national economies with one another. The result of such network might be ``contagion'': an ailing economy may severely damage creditors which, through financial exposure, can be affected by spiralling losses experienced by the debtor. For instance, a foreign investor may see their investments in foreign bonds significantly devalued if the bond price declines due to unexpected losses, or out-of-control debt management. The foreign investor's loss may be commensurate with the percentage investment in an ailing economy compared with their total financial exposure and reserves gained through income. Here size matters, of course, and 
the discussions about the magnitude of the Greek versus the Italian debt impacting on the Eurozone or world economy come to one's mind. The next example aims at illustrating how contagion effects can be modelled in the present asset pricing framework. We introduce a linear combination of cumulative random bridge processes defined by
\begin{equation*}
 \widetilde{L}^{(j)}_{tU}=\sum^n_{i} w^{(j)}_i\,L^{(i)}_{tU},
\end{equation*}
where $\{L^{(i)}_{tU}\}$ are increasing random bridges, e. g. gamma random bridges with joint terminal distribution and generated by independent gamma processes. The weight parameter $w^{(j)}_i$ measures the level of exposure to each cumulative process $\{L^{(i)}_{tU}\}$. Since the linear combination may not be the same for any creditor exposed to the pool $\{\widetilde{L}^{(j)}_{tU}\}$, further freedom is given through $j$-indexing the exposed entity (sovereign state, private company, etc.). For instance {\it Country A} may have a financial exposure of 15\% to {\it Country X} and 8\% to {\it Country Y}. The percentage exposures, which can be collected from various financial intelligence organizations, can be used to determine the weights for {\it Country A}. On the other hand, {\it Country B} may be exposed by 34\% to {\it Country Y}'s economic performance and by 65\% to {\it Country Z}'s. One sees that, even though {\it Country A} and {\it B} may not hold any of each others financial assets, they are linked to each other through 
the 
common 
exposure to {\it Counrty Y}'s economy, albeit to different levels. The bonds of {\it Country A} and {\it Country B} are expected to both show the impact of their respective exposures to {\it Country Y}. The bond price processes, 
\begin{equation}\label{bondcontagion}
P^{(j)}_{tT}=\frac{P^{(j)}_{0T}+b_j(T)\,A^{\mL\,(j)}_t}{P^{(j)}_{0t}+b_j(t)\,A^{\mL\,(j)}_t},
\end{equation}
of the exposed entity $j$ can be modelled, for example, along the lines of (\ref{debt-bond}) where
\begin{align}\label{GFSI}
&P^{(j)}_{0t}=\frac{f^{(j)}_0(t)+(U-t)^{n+1}f^{(j)}_1(t)}{1+U^{n+1}f^{(j)}_1(0)},& &b_j(t)=\frac{(U-t)^{n+1} f_1^{(j)}(t)}{1+U^{n+1}f^{(j)}_1(0)},&\nn\\ 
 &A^{\mL\,(j)}_t=\prod^n_{i=1}\left(1-w_i^{(j)}\right)^{m_i t}\exp\left(-a_j\,L^{(j)}_{tU}-\tfrac{1}{2}a_j^2 t+\widetilde{L}^{(j)}_{tU}\right)-1,& &0\le t\le T<U,&
\end{align}
and $0\le w_i^{(j)}\le 1$, $m_i>0$. Here, $\{L^{(j)}_{tU}\}$  is taken to be a Brownian random bridge associated with the ordinary budget of the entity $j$. The level of financial exposure to a specific economy can be measured, at least in part, in terms of debt instruments held. Hereafter, we look at a simulation of contagion due to exposure to sovereign debt of two foreign countries. In particular, we suppose that Germany and France are exposed to the Spanish and Italian economic environment. The level of exposures $w^{(j)}_i$, governing in part the levels of dependence among the debt holders, are selected as follows:
\begin{table}[H]
\begin{center}
  \begin{tabular}{@{} ccccc @{}}
   $w_i^{(j)}$ & GER & FRA & ESP & ITA \\ \\
    GER & 1 & 0 & 0 & 0 \\ 
    FRA & 0 & 1 & 0 & 0 \\ 
    ESP & 0.57 & 0.47 & 1 & 0 \\ 
    ITA  & 0.49 & 0.25 & 0 & 1 \\ 
  \end{tabular}
  \label{table:exposures}
  \end{center}
\end{table}
In this basic illustration, Germany has no exposure to the French economy, however it has exposure levels 0.57 and 0.49 to Spain and Italy, respectively. The numbers listed in the table are not normalised. Dependence among the four economies is also subject to the joint marginal distribution of the LRBs underlying the dynamics of the yield processes. The closer the time horizon $U$, the more the joint marginal distribution of the multivariate random variable $L_{UU}$ will govern the dynamics of the dependent yield processes. 
\paragraph{Simulation of contagion.} The scenario that follows is based on the contagion model (\ref{bondcontagion}) based on the debt exposure matrix $w^{(j)}_i$ above. The yield process $\{y^{(j)}_{tT}\}_{0\le t<T}$ and the spread process $\{s^{(j)}_{tT}\}_{0\le t<T}$ are defined by
\begin{align}
&y^{(j)}_{tT}=-\frac{\ln\left(P^{(j)}_{tT}\right)}{T-t},& &s_{tT}=-\frac{\ln\left(P^{(j)}_{tT}\right)}{T-t}+\frac{\ln\left(P^{GER}_{tT}\right)}{T-t},&
\end{align}
where country $j$'s spread is quoted against Germany's yield. The quantities in (\ref{GFSI}) are set as follows: The horizon of the financial market is $U=5$, and the maturity of bond $j$ is $T=2$. Germany's and France's bond are both exposed to Spain's and Italy's debt, thus $n=2$. Furthermore,
\begin{equation}\label{f0f1}
f_0^{(j)}(t)=\e^{-\alpha_j\,t},\qquad f_1^{(j)}(t)=\frac{\beta_j}{\ln(\gamma_j+t)} 
\end{equation}
where $\alpha_j$, $\beta_j$ and $\gamma_j$ are constants and chosen such that the conditions for $f^{(j)}_0(t)$ and $f^{(j)}_1(t)$ are satisfied, see (\ref{genpk}). For the simulation of the Brownian random bridge $\{L^{(j)}_{tU}\}$ we set for the information flow parameter $\sigma_j=0.5$, and the $X$-factor takes the values $\{0, 5, 6, 7\}$ in Germany's $\{0, 5, 8, 10\}$ and in France's case, and $\{8, 10, 15, 20\}$ in Italy's and Spain's case. In all cases, the {\it a priori} probabilities for the $X$-factor's outcome are the same, namely $\{0.7, 0.2, 0.05, 0.05\}$. For the simulation of the gamma random bridge $\{L^{(i)}_{tU}\}$, the $X$-factor takes the values $\{0,1,2,0\}$ in Germany's and $\{-4, -3, -2, 0\}$ in France's case, and $\{-4,-3,-2,0\}$ in Italy's and $\{-5, -4, -3, 0\}$ in Spain's case. The {\it a priori} probabilities for this $X$-factor's outcomes are $\{0.2,0.4,0.4,0\}$ for all four countries. 
Moreover, $a_j=0.5$ and $m_i=1$.
\begin{figure}[H]
\begin{center}
\includegraphics[scale=.31]{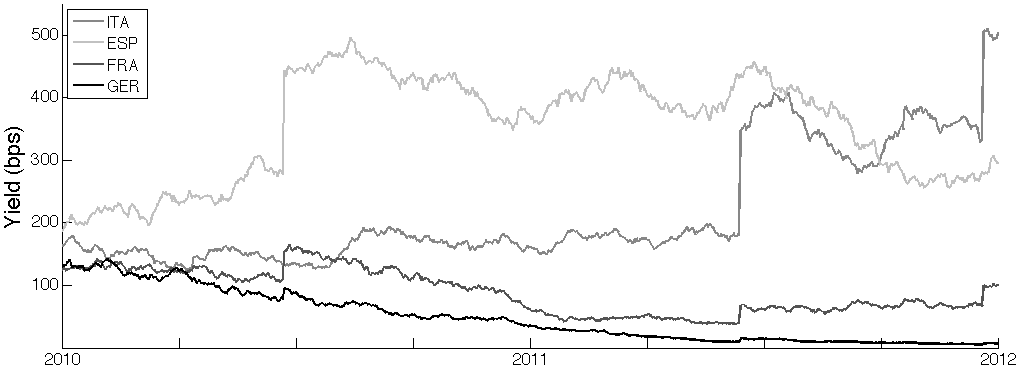}
\caption{Simulation of the yield process of the two-year-maturity bonds issued by Germany, France, Italy, and Spain.}
\vspace{-.25cm}
\label{figGBc1}
\end{center}
\end{figure}
In Figure \ref{figGBc1}, we see the impact of spiralling debt on the yield process of sovereign bonds. The significant jumps in the yield processes of Spain and Italy (first and second trajectories from above) are due to unexpected losses (e. g. bank bail-outs, natural disasters) over relatively short periods of time. These spikes then have repercussions on the yield processes of France and Germany to various degrees of severity. The level of repercussion on each exposed country is relative to the health of their own economy. In the simulation above, we see that France's yield process (third trajectory from above) needs more time than Germany's yield trajectory to recover from the shocks. The implication is that although Germany has a higher exposure level to Spain's and Italy's finances, it has a more robust domestic economy---with, e. g., higher growth rate---than France. Thus, Germany is in a position to better weather foreign economic shocks. Contagion effects, due to increased economic stress, are also 
observed in the 
behaviour of the spread process when comparing the performance of bonds issued by different sovereign states. 
\begin{figure}[H]
\begin{center}
\includegraphics[scale=.31]{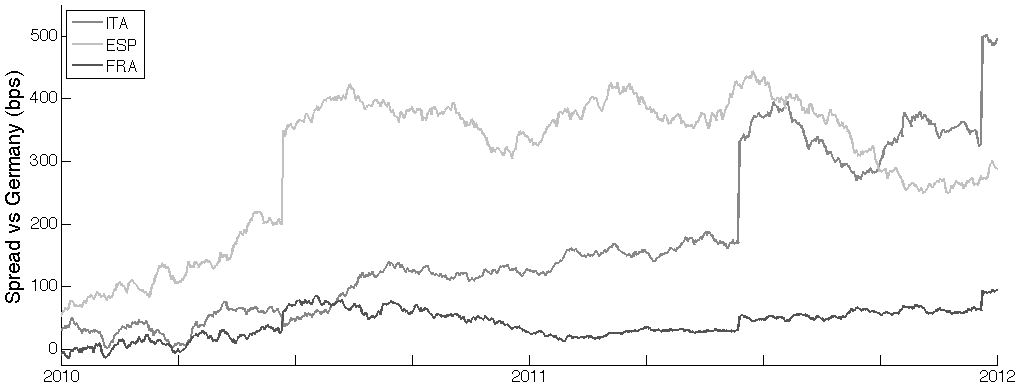}
\caption{Simulation of the price spread process for the two-year-maturity bonds issued by France, Italy, and Spain compared with the two-year bond issued by Germany.}
\vspace{-.25cm}
\label{figGBc2}
\end{center}
\end{figure}
In Figure \ref{figGBc2}, we plot the difference between the yield processes of France, Spain, and Italy when compared with the yield process of the German sovereign bond. We observe the widening of the spread level for Spain and Italy, due to the two upward jumps in economic deficit. France keeps the evolution of the spread between the yield of its bond and the one by Germany in check even though it takes the hit from the exposures to the Spanish and Italian economies.
\section{Conclusions}
Building on the work of Akahori \& Macrina \cite{am}, this article presents several novel contributions which include:
\begin{itemize}
\item a compact formalism for the weighted heat kernel pricing models in finite time with partial automatic calibration
\item characterisation of a class of versatile and tractable asset pricing models which includes the quadratic and exponential quadratic models constructed in reference \cite{am}
\item pricing formulae for caplets and swaptions with explicit examples
\item dynamical equations for price processes driven by continuous Markov processes and derivation of the endogenous interest rate and market price of risk models underlying the dynamics of bond prices 
\item transition to the risk-neutral measure consistent with the interest rate model, and identification of the source of model risk as a component of the risk premium
\item explicit, multi-factor, incomplete-market price models with jumps and potential application to portfolio assets across dependent market sectors 
\item generalised price models based on Fourier transforms and designed for weighted heat kernels driven by L\'evy random bridges 
\item extension of the weighted heat kernel approach to the pricing of general financial assets
\item explicit examples of price diffusions with stochastic discounting and stochastic volatility, and examples of price processes with heavy-tail distribution
\item modelling of contagion in financial markets and illustration of indirect credit-risk effects due to spiralling debt accumulation
\end{itemize}
The list of investigations connected with the asset pricing approach presented in this paper is by no means complete. The developed theory is applicable to the modelling of foreign exchange rates and the pricing of foreign exchange securities. Inflation-linked bonds and other indexed assets can be priced in a similar way. Forward rate agreements on LIBOR may be viewed as derivatives discounted by the short-term rate of interest, see Mercurio \cite{merc}. In this paper, the overnight rate of interest (e. g. EONIA) is modelled as part of the pricing kernel. How perceived credit risk and changes in liquidity may affect the dynamics of the pricing kernel, or how pricing kernel models could produce multi-curve discounting, is worth further research work. Then there are commodity assets, including energy and agricultural products, which may involve insurance contracts to hedge against substantial losses due to averse weather conditions. Insurance pricing models which include endogenous stochastic interest rate models are likely to be in line with hedge-instruments, e. g. interest rate swaps, traded in a financial market, and thus are more likely to be market-consistent (Solvency II).  In Section \ref{defic-mod}, effects on asset prices by perceived insolvency risk are considered in the dynamics of their returns. Further work in this direction, may include the modelling of ``costs of funding'' arising from shifts in asset prices due to the deterioration of a creditor's economic situation. Although fully-fledged credit risk models have not yet been developed within the present asset pricing framework (this is yet another project), the material in Section \ref{defic-mod} may also be looked at from the perspective of modelling ``credit valuation adjustments''. The bounds, within which the dynamics of certain price processes are confined, could be exploited to model the levels of sustainability for the costs of funding, or applied by regulators to impose time-dependent capital requirements. Although treated in this paper, the explicit modelling of dependence structures for asset portfolios remains somewhat in the background. The weighted heat kernel approach, however, offers a versatile basis for the inclusion of manageable dependence models that could be useful for the risk analysis of asset portfolios. One could begin with mean-variance optimisation whereby covariance matrices are explicitly modelled, and partial information about the underlying risk factors drives the portfolio. Another investigation may concern models for volatility surfaces, which arise from the selection of particular pricing kernels and their application to specific asset classes. Such an investigation extends to the analysis of the derived option price models and their calibration to data relevant for the pricing of assets. 
\\
%

\noindent {\bf Acknowledgments}.\,
The author is grateful to J.~Akahori, D.~Brigo, C.~Buzzi, S. Crep\'ey, M.~A.~Crisafi, D.~Filipovi\'c, M.~Forde, A.~E.~V.~Hoyle, M.~Kupper, C.~Mainberger, P.~McCloud, P.~A.~Parbhoo, M.~Riedle, J.~Sekine, D.~R.~Taylor for useful discussions, and to conference and seminar participants at the African Institute for Mathematical Science in Cape Town, the Technische Universit\"at Berlin, the Rand Merchant Bank in Johannesburg, the South African Reserve Bank in Pretoria, and the SIMC-2013 for helpful comments. Part of this work was carried out while the author was a member of the Department of Mathematics, King's College London.

\end{document}